\documentclass[longauth]{aa}  

\usepackage{graphicx}
\usepackage{txfonts}
\usepackage{subfig}
\usepackage{textgreek}
\usepackage{multirow}
\usepackage{mathtools}
\usepackage{stfloats}
\usepackage{color}
\usepackage{rotating}
\usepackage{enumitem}
\usepackage{natbib}
\usepackage{hyperref}
\usepackage{amsmath,amssymb}
\usepackage{epstopdf}

\bibpunct{(}{)}{;}{a}{}{,}

\newcommand{\FIG}[1]{}

\def\mso{\,{\rm M}_\odot}

\def\kms{\, {\rm km}\, {\rm s}^{-1}}

\def\msoy{\, \mso~{\rm yr}^{-1}}

\usepackage[normalem]{ulem}

\begin{document}

        \title{{\sc Atomium}: The astounding complexity of the near circumstellar environment of the M-type AGB star R Hydrae}
        
        \subtitle{I. Morpho-kinematical interpretation of CO and SiO emission}
        
        \author{Ward Homan
                \inst{1,2}
                \and
                Bannawit Pimpanuwat
                \inst{3,20}
                \and
                Fabrice Herpin
                \inst{10}
                \and
                Taissa Danilovich
                \inst{2}
                \and
                Iain McDonald
                \inst{3}
                \and
                Sofia H. J. Wallstr\"om
                \inst{2}
                \and
                Anita M. S. Richards
                \inst{3}
                \and
                Alain Baudry
                \inst{10}
                \and
                Raghvendra Sahai
                \inst{7}
                \and
                Tom J. Millar
                \inst{12}
                \and
                Alex de Koter
                \inst{5,2}
                \and
                C. A. Gottlieb
                \inst{17}
                \and   
                Pierre Kervella
                \inst{4}
                \and 
                Miguel Montarg\`{e}s
                \inst{4}
                \and
                Marie Van de Sande
                \inst{2}
                \and
                Leen Decin
                \inst{2,21}
                \and   
                Albert Zijlstra
                \inst{3}
                \and      
                Sandra Etoka
                \inst{3}
                \and
                Manali Jeste
                \inst{6}
                \and
                Holger S. P. Müller
                \inst{19}
                \and
                Silke Maes
                \inst{2}
                \and
                Jolien Malfait
                \inst{2}
                \and
                Karl Menten
                \inst{6}
                \and
                John Plane
                \inst{3}
                \and
                Kelvin Lee
                \inst{9}
                \and
                Rens Waters
                \inst{5}
                \and
                Ka Tat Wong
                \inst{11}
                \and
                Eric Lagadec
                \inst{13}
                \and
                David Gobrecht
                \inst{2}
                \and
                Jeremy Yates
                \inst{14}
                \and
                Daniel Price
                \inst{15}
                \and
                Emily Cannon
                \inst{2}
                \and
                Jan Bolte
                \inst{2}
                \and
                Frederik De Ceuster
                \inst{2,14}
                \and
                Joe Nuth
                \inst{16}
                \and
            Jan Philip Sindel
                \inst{22,2,20}
                \and
                Dylan Kee
                \inst{2}
                \and
                Malcolm D. Gray
                \inst{3,20}
                \and
                Ileyk El Mellah
                \inst{23,18,2}
        }
        
        \offprints{W. Homan}          
        
        \institute{
                $^{\rm 1}\ $Institut d’Astronomie et d’Astrophysique, Université Libre de Bruxelles (ULB), CP 226 -- 1060 Brussels, BE \\
            $^{\rm 2}\ $Institute of Astronomy, KU Leuven, Celestijnenlaan 200D B2401, 3001 Leuven, BE \\
                $^{\rm 3}\ $JBCA, Department Physics and Astronomy, University of Manchester, Manchester M13 9PL, UK \\   
                $^{\rm 4}\ $LESIA (CNRS UMR 8109), Observatoire de Paris, PSL, CNRS, UPMC, Univ. Paris-Diderot, FR \\   
                $^{\rm 5}\ $Astronomical Institute Anton Pannekoek, University of Amsterdam, Science Park 904, PO Box 94249, 1090 GE, Amsterdam, NL \\   
                $^{\rm 6}\ $Max-Planck-Institut für Radioastronomie, auf dem Hügel 69, 53121 Bonn, GE \\
                $^{\rm 7}\ $Jet Propulsion Laboratory, California Institute of Technology, 4800 Oak Grove Drive, Pasadena, CA 91109, USA \\
                $^{\rm 9}\ $Radio and Geoastronomy Division, Harvard-Smithsonian Center for Astrophysics, Cambridge, MA, USA \\
                $^{\rm 10}\ $Laboratoire d'astrophysique de Bordeaux, Université de Bordeaux, CNRS, B18N, Allée Geoffroy Saint-Hilaire, 33615 Pessac, FR. \\
                $^{\rm 11}\ $Institut de Radioastronomie Millimétrique, 300 rue de la Piscine, 38406 Saint Martin d’Hères, FR \\
                $^{\rm 12}\ $Astrophysics Research Centre, School of Mathematics and Physics, Queen’s University Belfast, University Road, Belfast BT7 1NN, UK. \\
                $^{\rm 13}\ $Laboratoire Lagrange, Université Côte d’Azur, Observatoire de la Côte d’Azur, CNRS, Boulevard de l’Observatoire, CS 34229, F-06304 Nice Cedex 4, FR. \\
                $^{\rm 14}\ $Department of Physics and Astronomy, University College London, Gower Street London, WC1E 6BT, UK \\
                $^{\rm 15}\ $School of Physics \& Astronomy, Monash University, Clayton, Vic 3800, AUS \\
                $^{\rm 16}\ $Solar System Exploration Division, Code 690 NASA’s Goddard Space Flight Center, Greenbelt MD 20771 USA \\
                $^{\rm 17}\ $Harvard-Smithsonian Center for Astrophysics, 60 Garden Street Cambridge, MA 02138, USA \\
                $^{\rm 18}\ $Center for Mathematical Plasma Astrophysics, Celestijnenlaan 200B, 3001 Leuven, BE \\
                $^{\rm 19}\ $I. Physikalisches Institut, Universität zu Köln, Zülpicher Str. 77, 50937 Köln, Germany \\
                $^{\rm 20}\ $National Astronomical Research Institute of Thailand, 260 Moo 4, T. Donkaew, A. Maerim, Chiangmai 50180, TH.\\
                $^{\rm 20}\ $School of Physics and Astronomy, University of St Andrews, North Haugh, St Andrews, Fife, KY16 9SS, UK.\\           
                $^{\rm 21}\ $School of Chemistry, University of Leeds, Leeds LS2 9JT, UK \\
                $^{\rm 22}\ $Centre for Exoplanet Science, University of St Andrews, North Haugh, St Andrews, KY169SS, UK.\\
                $^{\rm 23}\ $Institut de Planétologie et d’Astrophysique de Grenoble (IPAG), Université Grenoble Alpes, 38058 Grenoble Cedex 9, France
        }             
        
        \date{Received <date> / Accepted <date>}
        
        \abstract  
        {Evolved low- to intermediate-mass stars are known to shed their gaseous envelope into a large, dusty, molecule-rich circumstellar nebula which typically develops a high degree of structural complexity. Most of the large-scale, spatially correlated structures in the nebula are thought to originate from the interaction of the stellar wind with a companion. As part of the {\sc Atomium} large programme, we observed the M-type asymptotic giant branch (AGB) star R~Hydrae with the Atacama Large Millimeter/submillimeter Array (ALMA). The morphology of the inner wind of R~Hya, which has a known companion at $\sim$3500~au, was determined from maps of CO and SiO obtained at high angular resolution. A map of the CO emission reveals a multi-layered structure consisting of a large elliptical feature at an angular scale of $\sim$10'' that is oriented along the north--south axis. The wind morphology within the elliptical feature is dominated by two hollow bubbles. The bubbles are on opposite sides of the AGB star and lie along an axis with a position angle of $\sim$115$^\circ$. Both bubbles are offset from the central star, and their appearance in the SiO channel maps indicates that they might be shock waves travelling through the AGB wind. An estimate of the dynamical age of the bubbles yields an age of the order of 100 yr, which is in agreement with the previously proposed elapsed time since the star last underwent a thermal pulse. When the CO and SiO emission is examined on subarcsecond angular scales, there is evidence for an inclined, differentially rotating equatorial density enhancement, strongly suggesting the presence of a second nearby companion. The position angle  of the   major axis of this disc is $\sim$70$^\circ$ in the plane of the sky. We tentatively estimate that a lower limit on the mass of the nearby companion is $\sim$0.65~$\mso$ on the basis of the highest measured speeds in the disc and the location of its inner rim at $\sim$6~au from the AGB star.
    }
        
        
        \maketitle
        \section{Introduction} \label{intro}
        
        Stars with initial masses of between $\sim$0.8 and $\sim$8~$\mso$ evolve away from the main sequence (MS) and eventually become asymptotic giant branch (AGB) stars,
         characterised by considerable mass-loss of $\sim$10$^{\rm -8}$ to a few times $\sim$10$^{\rm -5}$~$\msoy$. For mass-loss rates above $\sim$10$^{\rm -7}$~$\msoy$, this process operates on timescales smaller than the nuclear burning timescale, and controls the further evolution of the star \citep[see e.g. first footnote in ][]{McDonald2018}.
        
        Mass loss is thought to be a two-step process. First, stellar surface pulsations \citep[][and references therein]{McDonald2018} bring matter in the stellar atmosphere to regions favourable for the formation of dust \citep{Bowen1988,Hofner2018}. In the second step, the dust that is thought to be primarily formed in post-shock regions \citep{Freytag2017} couples efficiently with the stellar radiation field, is accelerated outwards, and drags the surrounding gas along \citep{Gilman1972}. The detection of highly complex morphological structures in the inner wind of AGB stars at high spatial resolution \citep[e.g.][ and references therein]{Decin2020} revealed that the mechanisms shaping these dusty winds are often dominated by binary interactions.
        
        The observation of non-spherical distributions of matter of the evolutionary progeny of the AGB stars, the post-AGB and planetary nebula systems \citep{Sahai2007,DeMarco2009,Lagadec2011,Sahai2011}, suggests that this morphological transition develops from the small-scale structure on the AGB \citep{Decin2020}. In order to shed light on the specific mechanisms that shepherd the morphological evolution of evolved stars, it is essential to map and classify the wind morphologies of the circumstellar environments (CSEs) of AGB stars.
        
        It is in this context that we present a morphological analysis of the CSE of the M-type AGB star R Hydrae. R~Hya is a Mira-type pulsator with a period of 385 days \citep{Zijlstra2002}. However, before settling to this period of 385 days in the year 1950, the pulsation period of R~Hya had slowly decreased at a rate of $\sim$50 days per century, starting from 495 days in the year 1770, although the uncertainties on the oldest measurements are substantial \citep{Zijlstra2002}. The detection of technetium in its stellar photosphere indicates that the star is on the thermally pulsing AGB track \citep{Little1987}, though the detection is very weak \citep{Uttenthaler2011}. 
        
        
        \emph{Hipparcos} estimates a distance to R Hya of 124$^{+12}_{-10}$ pc \citep{VanLeeuwen2007}, and while \emph{Gaia} DR2 estimated a much larger distance (239$^{+91}_{-53}$ pc, \citet{BailerJones2018}), \emph{Gaia} EDR3 returns the estimate to 148$^{+11}_{-10}$ pc. However, the first estimate of its distance comes from its space motion and photometry \citep{Eggen1985}. These latter authors put R~Hya at $\sim$165 pc, which we adopt in this work to provide consistency with previous studies.
        
        R~Hya has a 3$\sigma$ proper motion discrepancy between \emph{Hipparcos} (--57.7 $\pm$ 0.9, 12.9 $\pm$ 0.5 mas yr$^{-1}$) and \emph{Gaia} DR3 (--54.2 $\pm$ 0.5, 11.8 $\pm$ 0.3 mas yr$^{-1}$). While this may be due to astrometric noise, it can also be interpreted as a signature of binarity \citep{Kervella2019}. R~Hya has a twelfth-magnitude companion, \emph{Gaia} EDR3 6195030801634430336, at an angular separation of $\sim$21$^{\prime\prime}$ \citep{Mason2001}. It has an almost identical proper motion vector to R Hya and a similar parallax (\emph{Gaia} EDR3: --55.68 $\pm$ 0.21, 13.28 $\pm$ 0.14 mas yr$^{-1}$; 128.4$^{+3.3}_{-3.1}$ pc), and therefore likely represents a true binary association. \citet{Anders2019} estimates its mass at $\sim$0.8 M$_\odot$. However, this companion is too distant to cause the proper motion discrepancy in R Hya, raising the possibility that it is a triple system.
        
        \citet{Eggen1985} suggested that R~Hya is a member of the Hyades supercluster, which allowed him to estimate its age to be $\sim$5--10 $\times$10$^{8}$~yr, and its initial mass to be $\sim$2~$\mso$. This estimate was further refined by \citet{Decin2020}, who used the $^{\rm 17}$O to $^{\rm 18}$O isotope ratio of R~Hya \citep{Hinkle2016} to deduce an initial mass \citep{DeNutte2017} of $\sim$1.35~$\mso$. \citet{Haniff1995} estimated the angular diameter of R~Hya to be approximately 0.030'' in the near-infrared. The optical diameter of the star has been measured to be 0.023'' \citep{Richichi2005}. \citet{Haniff1995} also used the bolometric flux to derive a stellar surface temperature of $\sim$2650~K, whereas \citet{Feast1996} used J-K relations to derive a temperature of $\sim$2800~K. The latest surface temperature estimate was performed by \citet{DeBeck2010}, who used the V-K colour to deduce a much lower effective temperature of $\sim$2100~K. 
        
        The CSE of R~Hya has been the focus of many earlier studies. The latest radiative transfer modelling of CO lines by \citet{Schoier2013} yielded a mass-loss rate of 2.1$\times$10$^{\rm -7}$~$\msoy$. \citet{Hashimoto1998} used IRAS data to show that R~Hya possesses an extended dust envelope composed primarily of silicate dust. With no evidence of hot dust in the interior of the envelope, these latter authors suggested that the system may have halted its mass loss over the past century. \citet{Decin2008} traced the mass-loss history of R~Hya using the ISO spectrum and modelled a sequence of ro-vibrationally excited CO lines. They report a change in mass-loss rate $\sim$230 years ago, from $\sim$3$\times$10$^{\rm -7}$~$\msoy$ to a current rate that is almost 20 times lower. These findings are in agreement with the predictions made by \citet{Zijlstra2002} based on the decline of the pulsational period, and are believed to be associated with the occurrence of a recent thermal pulse \citep[TP; ][]{Iben1983,Habing2004}. 
        
        \citet{Ueta2006} used \emph{the Spitzer space telecope} to spatially resolve the bow shock of R~Hya, which is located at a distance of $\sim$100'' to the northwest of the AGB star, and was later modelled by \citet{Wareing2006} using a grid-based hydrodynamics code. \citet{Cox2012} imaged the bow shock of R~Hya with the PACS instrument on the Herschel space telescope. The inner regions of the star were also probed with the \emph{MIDI/VLTI} instrument by \citet{ZhaoGeisler2012} to investigate the distribution of dust. R~Hya is also part of the DEATHSTAR project\footnote{https://www.astro.uu.se/deathstar/index.html}, within which it has been observed with the ALMA compact array in Band 6 at a resolution of $\sim$5'' \citep{Ramstedt2020}, revealing only a barely resolved signal. Besides these studies, no spatially resolved maps of the nebular environment near to the AGB star R~Hya are found in the literature.
        
        As part of the {\sc atomium} ALMA large programme (2018.1.00659.L., PI L. Decin), 
        we conducted the first spatially resolved Band~6 observations of the CSE of R~Hya with three different antenna configurations. The {\sc atomium} large programme consists of a homogeneous 27~GHz-wide spatially resolved  spectral line survey of 17 Asymptotic Giant Branch (AGB) and Red Supergiant (RSG) stars \citep[see][for an overview see Gottlieb et al. 2020, \emph{subm.}]{Decin2020}. Among the eight~oxygen-rich semi-regular and Mira variables in the {\sc atomium} sample, the most molecule-rich star is R~Hya (Wallstr{\"o}m et al., \emph{in prep.}). In the present paper we investigate the morphology of R~Hya's wind, as revealed by the CO 
        and SiO emission. These types of detailed morphological descriptions provide the fundamental framework for a quantitative understanding of the mass-loss phenomena that lead to the demise of most stars in the Universe but are still poorly understood. 
        
        The paper is organised as follows. We present a summary of the technical aspects of the observations, and the reduction procedure in Sect. \ref{obs}. In Sect. \ref{continuum}, we investigate the continuum emission, which is followed by an analysis of the CO
         and SiO emission in Sect. \ref{molobs}. Our findings are discussed in Sect. \ref{discus}, and are summarised in Sect. \ref{summ}.
        
        \section{Data acquisition and reduction} \label{obs}
        
    As part of the {\sc atomium} ALMA large programme, R~Hya was observed with three different ALMA antenna array configurations: compact, mid, and extended, with baseline ranges of 15-500m, 15-1398m, and 111-13894m, respectively. The compact-configuration data were obtained in the period from 27 December 2018 to 6 January 2019, the mid-configuration data from 25 to 27 October 2018, and the extended-configuration data from 9 to 12 July 2019. The data from all three were combined to produce the data shown below.
    
    At the epoch of the extended configuration observations, 9 to 12 July 2019,
    the peak of a 2D Gaussian fitted to the continuum peak was at ICRS 13:29:42.70212  -23.16.52.5146. To combine the images, all epochs were aligned with this position because there is significant proper motion between observations in other configurations. The position accuracy is approximately 2 milliarcseconds, mainly due to phase solution transfer errors, and the flux scale is accurate to about 10\%. The spectral coverage consisted of 16 spectral windows totalling 27.1875~GHz between 211 and 275~GHz (only half the spectral windows were observed in the compact configuration); see Fig. S1 in \citet{Decin2020}, and Tables 2 and E.3, and Fig. 1 in Gottlieb et al. (\emph{submitted}), which we refer to for more details on the observations and data processing. The synthesised beam size and the maximum recoverable scale (MRS) of the continuum are provided in Table \ref{GENERAL}. These values are also typical of the line data cubes in each configuration. The procedure for constructing the final cubes of CO and SiO is identical to what is described by \citet{Homan2020b},  with a few minor variations, notably in weighting (see following paragraph) during imaging to produce the desired resolutions as given in Table \ref{LINE}.
    
    In imaging the combined data, giving equal weight to all baselines provides the highest resolution for emission with a high-enough surface brightness, but is more prone to clean instabilities. A $uv$ plane Gaussian taper of FWHM 0.03'' (equivalent to a baseline of $\sim$9~km at 1.3~mm wavelength) was applied, that is, the longest baselines were smoothly weighted down, resulting in the beams given in Tables \ref{GENERAL} and \ref{LINE}.  For CO, an additional cube with a larger taper of 0.15'' (equivalent to ~1.8 km) was made to improve sensitivity to factor-seven-lower surface brightness extended emission. This came at a price of decreased resolution by a factor of three to four. We therefore henceforth refer to these cubes as the high-resolution (HR) and low-resolution (LR) CO cubes. This was not necessary for the SiO cube
    because its emission is confined to the vicinity of the AGB star (see Sect. \ref{SiOobs}). 
    Spectra were also extracted using different circular apertures centred on the star, whose diameters are given in the relevant figures.
        
        \begin{table}
                \caption{Summary of general and continuum data specifications, computed at 250~GHz.}
                \centering          
                \label{GENERAL}
                \begin{tabular}{llccl}
                        \hline\hline
                        \noalign{\smallskip}
                        Config. & Beam  & \multicolumn{2}{c}{Continuum} & MRS \\
                        & (mas$\times$mas,$^\circ$) & peak & rms & ('')\\
                        \noalign{\smallskip}
                        \hline    
                        \noalign{\smallskip}
                        extended & 34$\times$25, 67    & 41.86 & 0.27 & 0.6 \\
                        \noalign{\smallskip}
                        mid      & 256$\times$233, 70  & 54.44 & 0.10 & 3.5 \\
                        \noalign{\smallskip}
                        compact  & 830$\times$600, 79  & 65.55 & 0.06 & 8.7 \\
                        \noalign{\smallskip}
                        combined & 57$\times$47, 63    & 51.19 & 0.035 & 8.7 \\
                        \hline
                \end{tabular}
                        \vspace{1ex}\\
                        \begin{flushleft}
                        \textbf{Note}: The third number in the beam column represents the position angle of the major axis of the beam. mas = milliarcseconds. Peak flux density is expressed in mJy/beam, and rms flux density expressed in mJy. MRS stands for maximum recoverable scale.
                        \end{flushleft}
        \end{table}
        
        \begin{table*}
                \caption{Summary of the line cube data specifications.}
                \centering          
                \label{LINE}
                \begin{tabular}{lllllll}
                        \hline\hline
                        \noalign{\smallskip}
                        Molecule & Transition & Rest freq. & Taper & Chan. width & Channel noise rms               & Combined beam            \\
                        &            & (GHz)     & ('') & ($\kms$)    & ext-mid-comp-comb (mJy/beam) & [mas$\times$mas,PA ($^\circ$)] \\
                        \noalign{\smallskip}
                        \hline    
                        \noalign{\smallskip}
                        $^{12}$CO (HR) & $J$=2$-$1 & 230.5380 & 0.03 & 1.3 & 0.9\ --\ 1.8\ --\ 2.7\ \ \ --\ 0.6 & $87\times67$, 70 \\
                        \noalign{\smallskip}
                        $^{12}$CO (LR) & $J$=2$-$1 & 230.5380 & 0.15 & 1.3 & 0.9\ --\ 1.8\ --\ 2.7\ \ \ --\ 0.9 & $242\times255$, 72 \\
                        \noalign{\smallskip}
                        SiO & $J$=5$-$4 & 217.1050 & 0.03 & 1.35 & 1.0\ --\ 2.1\ --\ 3.0\ \ \ --\ 0.7 & $88\times66$, 70 \\
                        \hline
                \end{tabular}
                \vspace{1ex}\\
                \begin{flushleft}
                        \textbf{Note}: All lines are in $\varv = 0$. HR and LR refer to high- and low-resolution cubes; see Sect. \ref{obs}. ext = extended array data, mid = mid-sized array data, comp = compact array data, comb = combined data.
                \end{flushleft}
        \end{table*}

        \begin{figure}[]
                \centering
                \includegraphics[width=9cm]{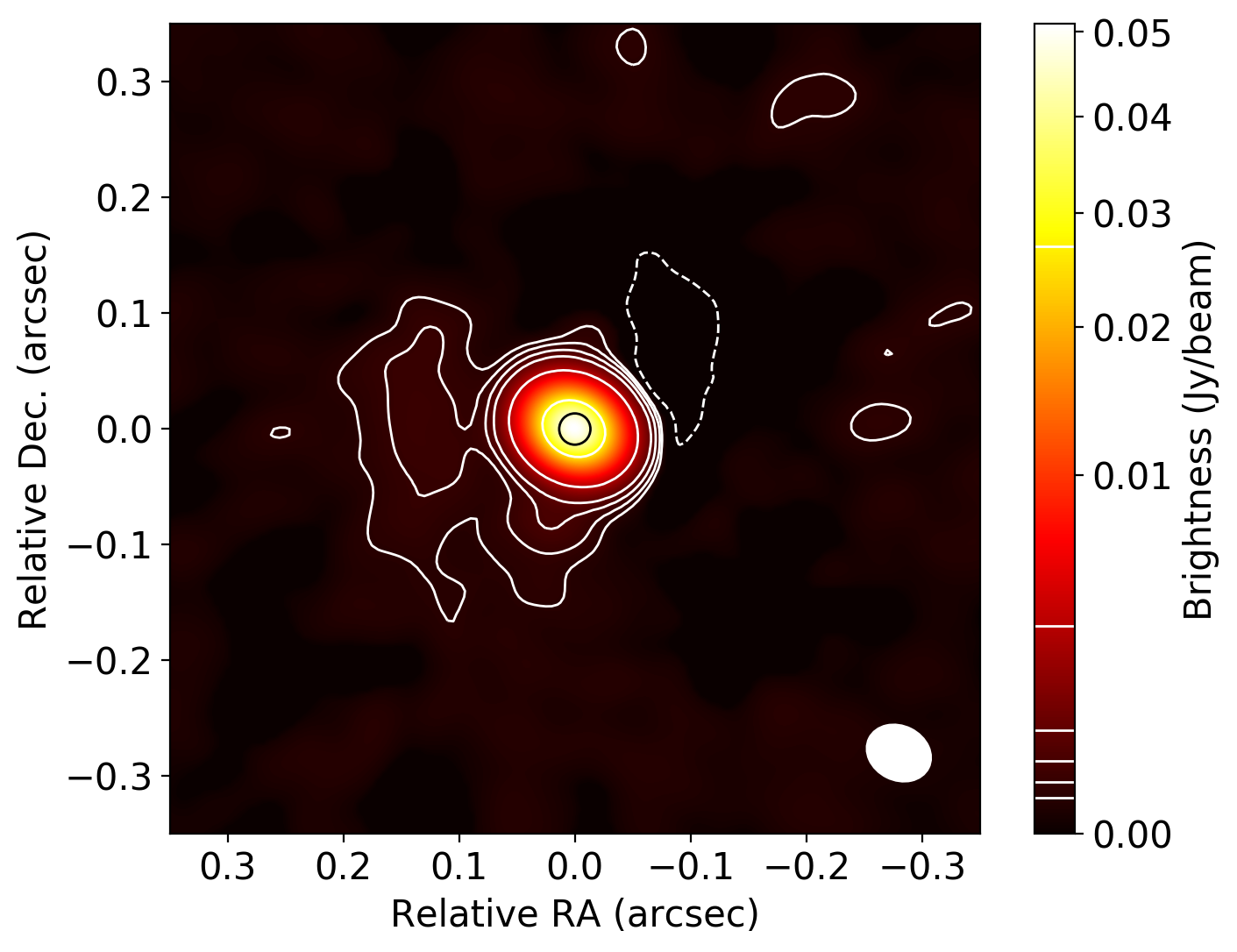}
                \caption{Continuum emission of the combined array data of R~Hya, at an effective mean frequency of 250~GHz. Contours are drawn at -3, 3, 6, 12, 24, 96 and 768 times the continuum rms noise value (3.5 $\times {\rm 10}^{\rm -5}$ Jy/beam), negative contours are dashed. The ALMA beam size (0.057''$\times$0.047'') is shown in the bottom right corner. The stellar photosphere as determined from the visibility fitting is drawn as a small black circle in the centre.}
                \label{cont}
        \end{figure}

        \begin{figure}[]
                \centering
                \includegraphics[width=9cm]{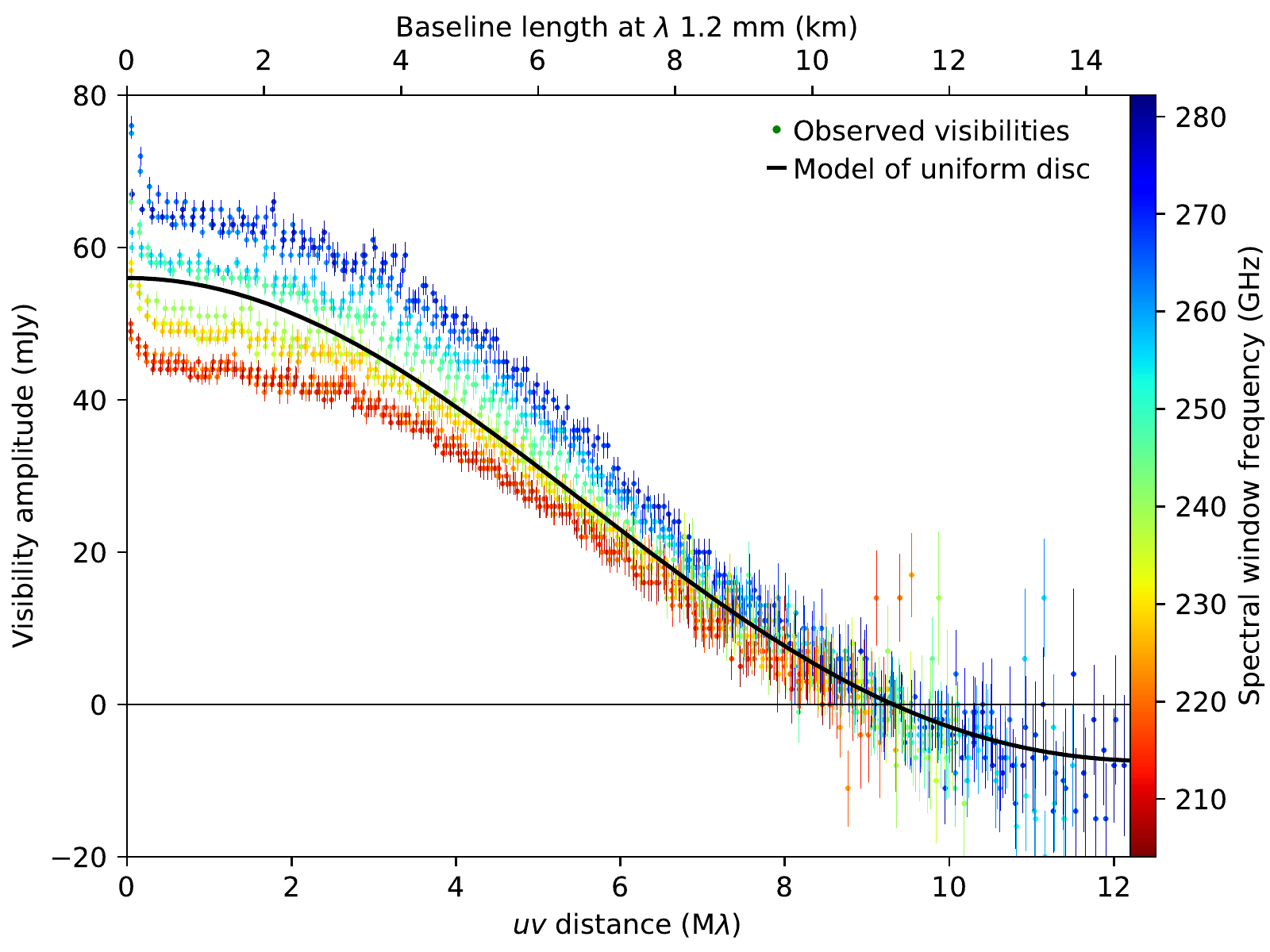}
                \caption{Continuum observed visibility amplitudes (plotted in colour for
                        each of the 16 spectral windows) as a function of uv distance and the corresponding baseline length. The black line shows the fitted model for a uniform disc of radius 13.5 milliarcseconds and flux density 56~mJy.  See text for more details.}
                \label{uvfit}
        \end{figure}
        
        \section{ALMA continuum emission} \label{continuum}
        
        The continuum images were made from line-free channels comprising 23.55, 19.02, and 10.09~GHz in the extended, mid, and compact configurations, respectively, spread over 211-275~GHz. The effective mean central frequency is 250 GHz. In Fig. \ref{cont} we show the continuum emission of the combined dataset, which consists of one bright, centrally condensed, slightly resolved patch of emission. Its total flux density within the 3$\times \sigma_{\rm rms}$ contour is $\sim$56~mJy, mainly dominated by the flux density within the 12$\times \sigma_{\rm rms}$ contour. The slight elongation is due to the beam shape and the exact distribution below $\sim$12$\times \sigma_{\rm rms}$ is unreliable, as there is a weak negative artifact opposite the eastern elongation. This is probably due to residual phase errors in the compact configuration data and its lower position accuracy. Nonetheless, in the combined image, there is a total flux density of 68 mJy within a 1'' radius of the star, close to the specific intensity peak of 65.55~mJy/beam detected at the largest resolution. Outside this radius (but within $\sim$90\% primary beam sensitivity), the combined map noise has a Gaussian distribution and sums to very close to zero as expected for an interferometry image.
        
        Figure \ref{uvfit} shows the calibrated visibility amplitudes averaged into 100 bins over the uv distance range for each of the 16 spectral windows (see Gottlieb et al. \emph{subm}, Tables 2 and E.3). These are shown by symbols colour-coded by frequency. The error bars are the standard deviations divided by four, as appropriate for comparison with a fit to the average of all 16 spectral windows. The scatter on the longest baselines is due to some combination of increased standard deviations due to fewer samples per bin; increased phase noise; and possibly the small-scale structure of the star. The shape of the combined continuum visibilities in Fig. \ref{uvfit} strongly suggests that it originates from an object that appears as a uniform disc in the plane of the sky. Therefore, we fitted them with a uniform disc model using the CASA task uvmodelfit.
        
        The black curve in Fig. \ref{uvfit} represents the Fourier-transform of a uniform disc of diameter 0.027'' ($\sim$4.5~au) and flux density of 56~mJy computed at the intermediate frequency of 250~GHz (1.2~mm). The disc model is an excellent fit to the visibilities. Its flux density corresponds to a brightness temperature $T_B$ of $\sim$2500~K, which was determined using the relation
        \begin{equation}
        T_B = \frac{1.97 \times 10^{-8} S_\nu c^2}{\nu^2 \theta_{maj}^2} {\rm K},
        \end{equation}  
        where $S_\nu$ is the source flux density in mJy, $c$ is the speed of light in $m s^-1$, $\nu$ is the observation frequency in GHz, and $\theta_{maj}^2$ is the diameter of the disc in milliarcseconds. This temperature is consistent with photospheric temperatures of R~Hya presented in the literature \citep{Haniff1995,Feast1996,DeBeck2010}. In addition, the diameter of the fitted disc around $\lambda$ = 1.2~mm is 17\% greater than the optical diameter of the star, which was measured to be 0.023'' by \citet{Richichi2005}. This difference is comparable to measurements by \citet{Vlemmings2019} of four other AGB stars, for which the millimeter-wave diameter is $\sim$13\% to 55\% bigger than the optical diameter because the temperature at which the optical depth approaches unity decreases with wavelength and therefore the apparent diameter increases. The spectral spread of the emission, as shown in Fig. \ref{cont}, corresponds to a spectral index of  between 1.8 and 1.9 (see Fig. \ref{specIn}). This is consistent with the imaging spectral index of the star derived from our multi-frequency imaging. Combined, these are strongly indicative that the disc model likely represents the stellar photosphere.
 
        The formal errors reported by the fitting task are known to be underestimated, but given the coincidence with the flux density inside the 12$\times \sigma_{\rm rms}$ contour we estimate the size error by analogy with Gaussian fitting as ($\sqrt{\rm 2}$/2)*(beam/12) = 0.003''. There is 5\% uncertainty in the flux density of ALMA standard calibrators. Additional amplitude errors due to imperfect calibration give a total uncertainty of typically 7\% per data set. Table \ref{GENERAL} shows that there is a consistent increase in flux density with measuring area which suggests that the overall flux density scale accuracy of the combined data could be as good as $\sim$10\%. 

        The visibilities show a small excess of about 8~mJy on the shortest baselines $<\sim$400 kilo-lambda, or $>\sim$0.5'' radius from the star. This is also seen from the compact configuration which detected ~10 mJy more than in smaller beams (see Table \ref{GENERAL}). This is likely to be emission from warm dust, but the signal-to-noise ratio of the extended emission is too low to measure its spectral index. \citet{Dehaes2007} measured $\sim$94.7~mJy within 30'' of the star, at 1.2 mm wavelength, similar to our observations. Assuming that the star contributes 56~mJy, this leaves $\sim$39~mJy dust emission, of which we are losing $\sim$30~mJy or about 75\% of the dust flux. This is to be expected, as the largest ALMA synthesised beam is <0.0005 of a 30'' diameter area.  Thus, the average surface brightness of the dust is <0.02~mJy/beam for our beam sizes, which is below the noise threshold. In addition, we are not sensitive to smooth emission on scales greater than $\sim$9''. We therefore probably only detect the warmest, least diffuse dust within $\sim$10 stellar radii (see Fig. \ref{SED}). As a consequence, the ALMA continuum is likely dominated by photospheric emission.

    \begin{figure}[]
        \centering
        \includegraphics[width=9cm]{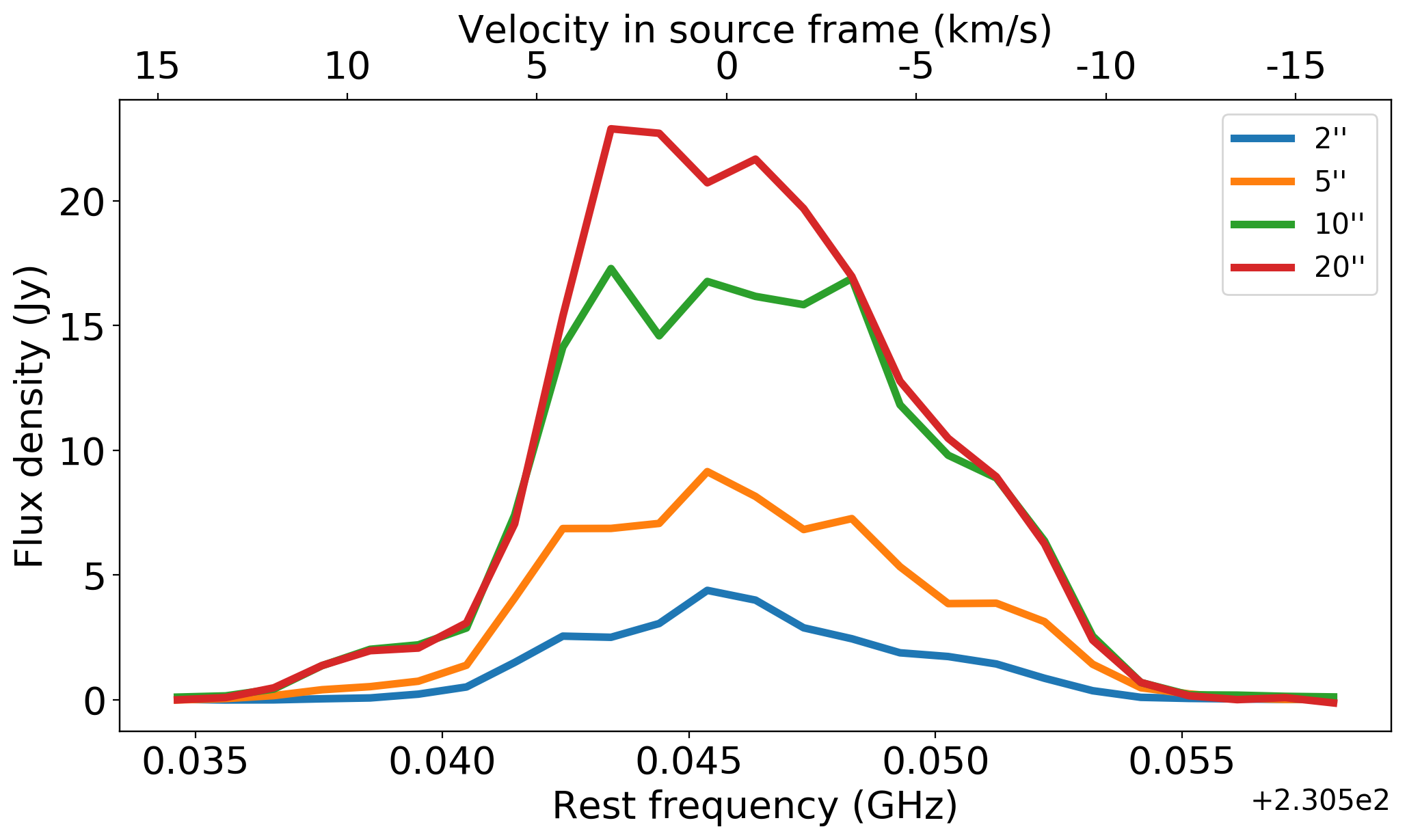}
        \caption{Specrum of the CO v=0 $J=\ $2$-$1, LR cube, obtained for different extraction aperture diameters. The frequency axis is adjusted to the stellar velocity. The scale bar at the top marks the LSR velocity of the star.}
        \label{COline}
    \end{figure}

    \begin{figure*}[]
        \centering
        \includegraphics[width=17cm]{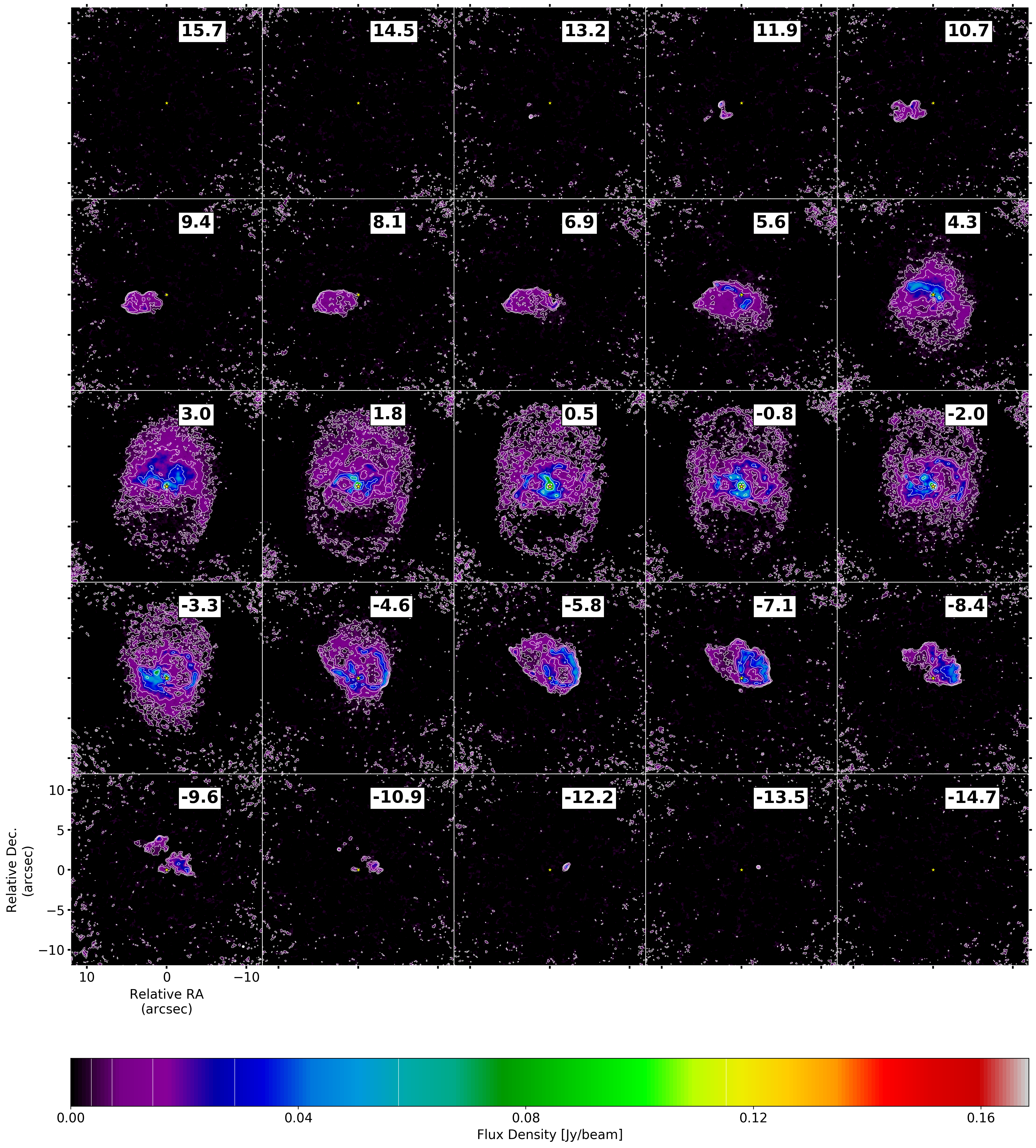}
        \caption{Channel maps showing the spatially resolved emission of the CO v=0 $J=\ $2$-$1 line in the $\pm \sim$15~$\kms$ velocity range. The maps were constructed using a taper weighting of 0.15''. $\sigma_{\rm rms}$ = 0.9$\times {\rm 10}^{\rm -3}$ Jy/beam, beam size =  (0.234''$\times$0.225'')'. The velocities have been corrected for $v_{*}$=-10.1~$\kms$. Contours are drawn at 3, 6, 12, 24, 48, and 96$\times \sigma_{\rm rms}$. Angular scales are indicated in the bottom left panel. The maps are centred on the continuum peak at position (0,0), and are indicated by a small yellow star.}
        \label{COchan1}
    \end{figure*}

    \begin{figure*}[]
        \centering
        \includegraphics[width=17cm]{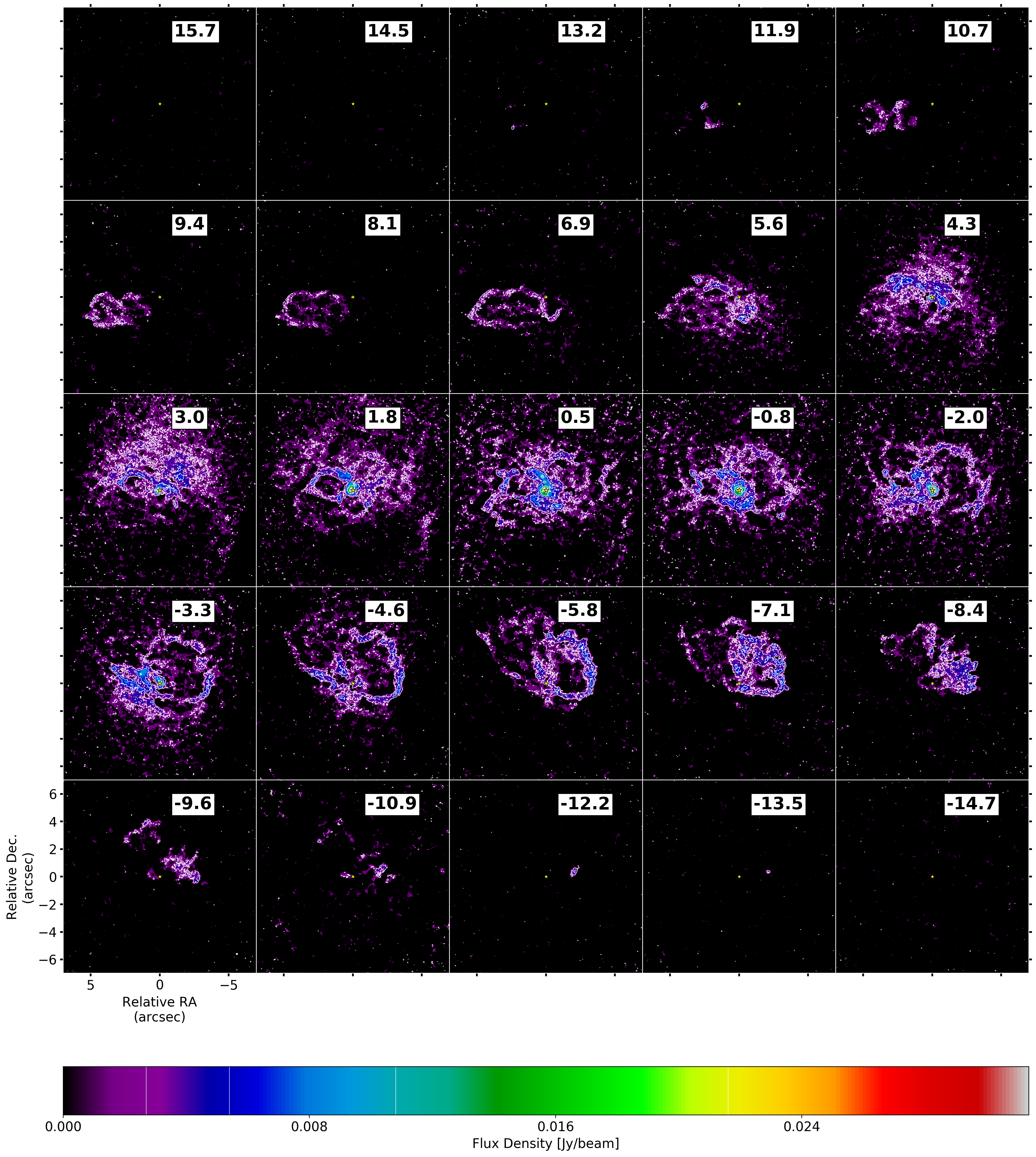}
        \caption{Same as Fig. \ref{COchan1}, but for the combined array data constructed using a taper weighting of 0.03''. $\sigma_{\rm rms}$ = 0.6$\times {\rm 10}^{\rm -3}$ Jy/beam, beam size = (0.087''$\times$0.067'').}
        \label{COchan2}
    \end{figure*}
        
        \section{ALMA molecular emission} \label{molobs}
        
        In this section we provide a comprehensive overview of the morphological features identified in the spatially resolved emission maps of the CO v=0 $J=\ $2$-$1
        \citep[230.5358 GHz, ][]{1997JMoSp.184..468W}
         and SiO v=0 $J=\ $5$-$4 
        \citep[217.1050 GHz, ][]{2013JPCA..11713843M} 
        molecular lines. Each of these molecules provides a specific diagnostic of the wind morphology. The CO molecule is primarily collisionally excited, making it an excellent tracer of density. However, there exists the possibility of radiative excitation of the v=1 vibrational band (and ensuing downward cascade) via 4.6 micron emission, and so prudence is recommended when interpreting CO emission that coincides with particularly dusty environments.        The high bond energy of SiO makes it a stable and easily formed product of local thermodynamical equilibrium (LTE) chemistry \citep{Gail2013b}. It is therefore highly abundant in the inner wind of M-type AGB stars. In combination with the high Einstein coefficient of the lower rotational transitions (in comparison with CO), these spectral lines yield high signal-to-noise ratios in the inner wind, which is why they  are considered an excellent tracer of the dynamics of the gas near the AGB star \citep{Kervella2016,Homan2018b,Decin2020}. 
        
        For optimal appraisal of the fine details in all maps throughout the paper, we recommend viewing all images on screen. For all maps, north is up, and east is left. All velocities in this paper are expressed with respect to the systemic velocity of $\sim\!{\rm -}\!{\rm 10.1}$~$\kms$ \citep{Danilovic2015}.
        
        \begin{figure*}[]
                \centering
                \includegraphics[width=13.5cm]{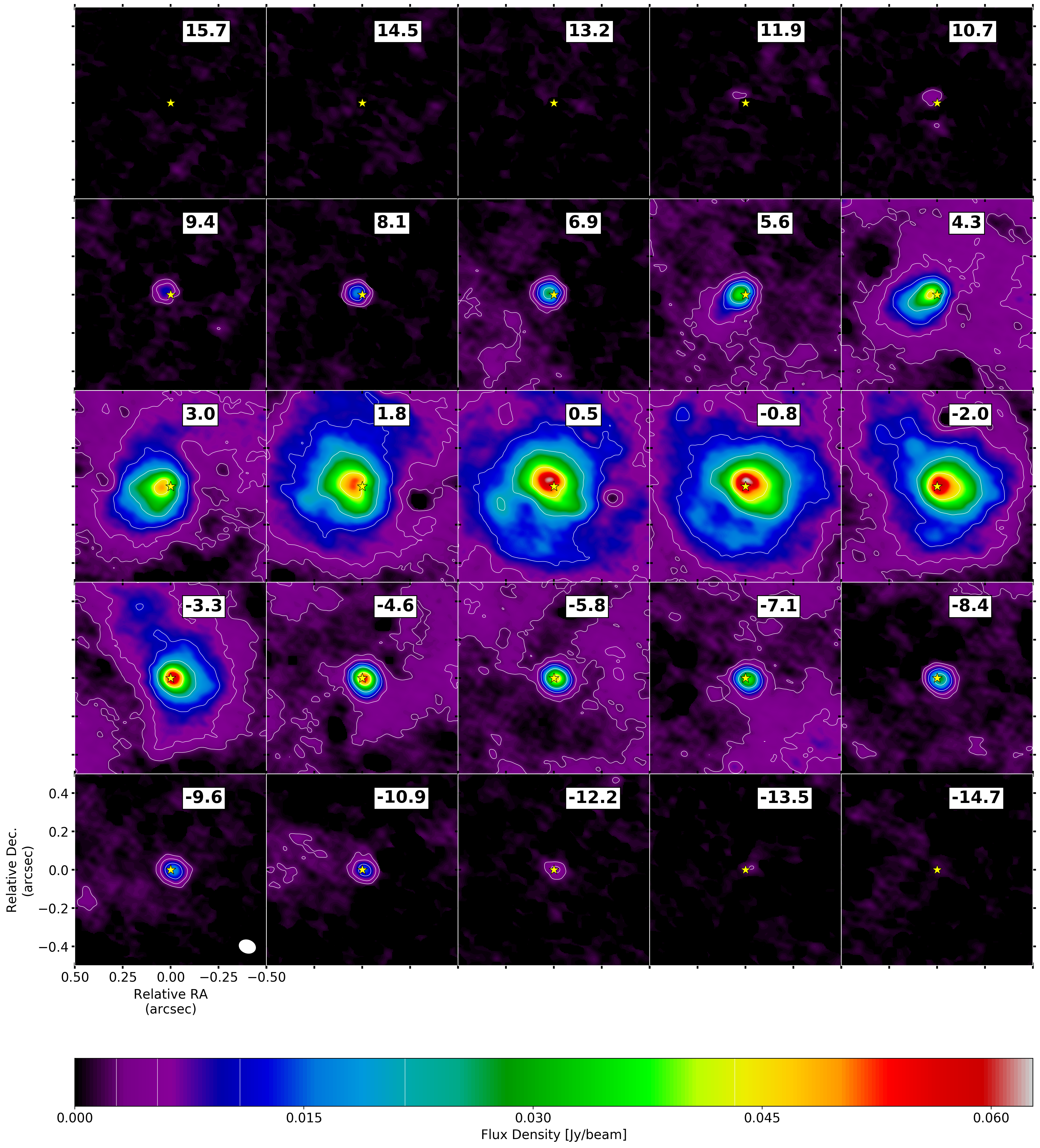}
                \caption{Same as Fig. \ref{COchan2}, but zoomed-in on the inner 1''.}
                \label{COchanZ}
        \end{figure*}

        \begin{figure}[htp!]
                \centering
                \includegraphics[width=8.5cm]{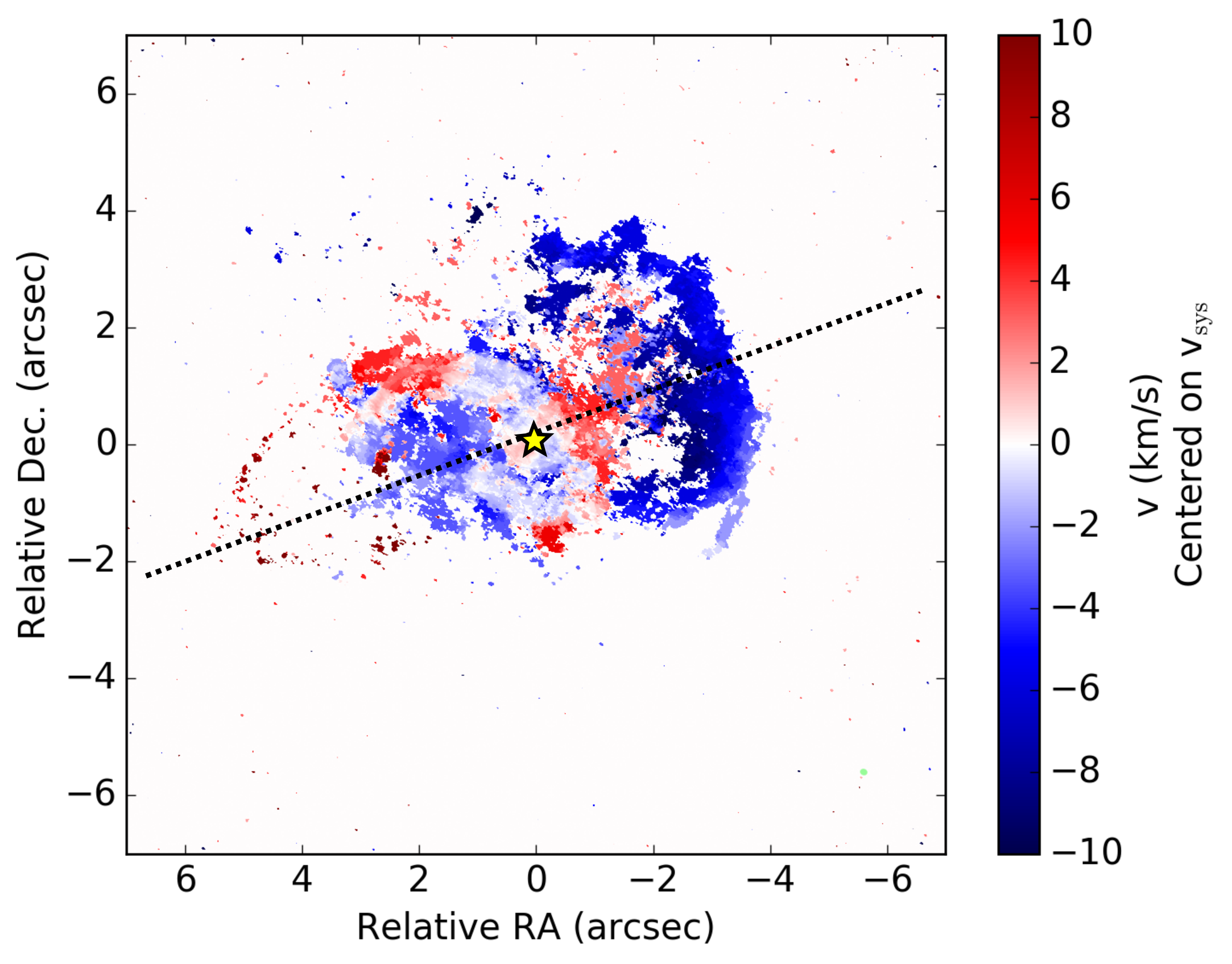}           \caption{Moment 1 map of the HR CO data. The dashed line represents the alignment
                        axis separating the eastern and western bubbles, which has a position angle of $\sim$115$^\circ$ (see also \emph{orange} line in Fig. \ref{COpvd}). The continuum peak is indicated by a yellow star. The small green ellipse in the bottom right corner is the ALMA beam.}
                \label{COmom1}
        \end{figure}

    \begin{figure*}
        \centering
        \includegraphics[width=18cm]{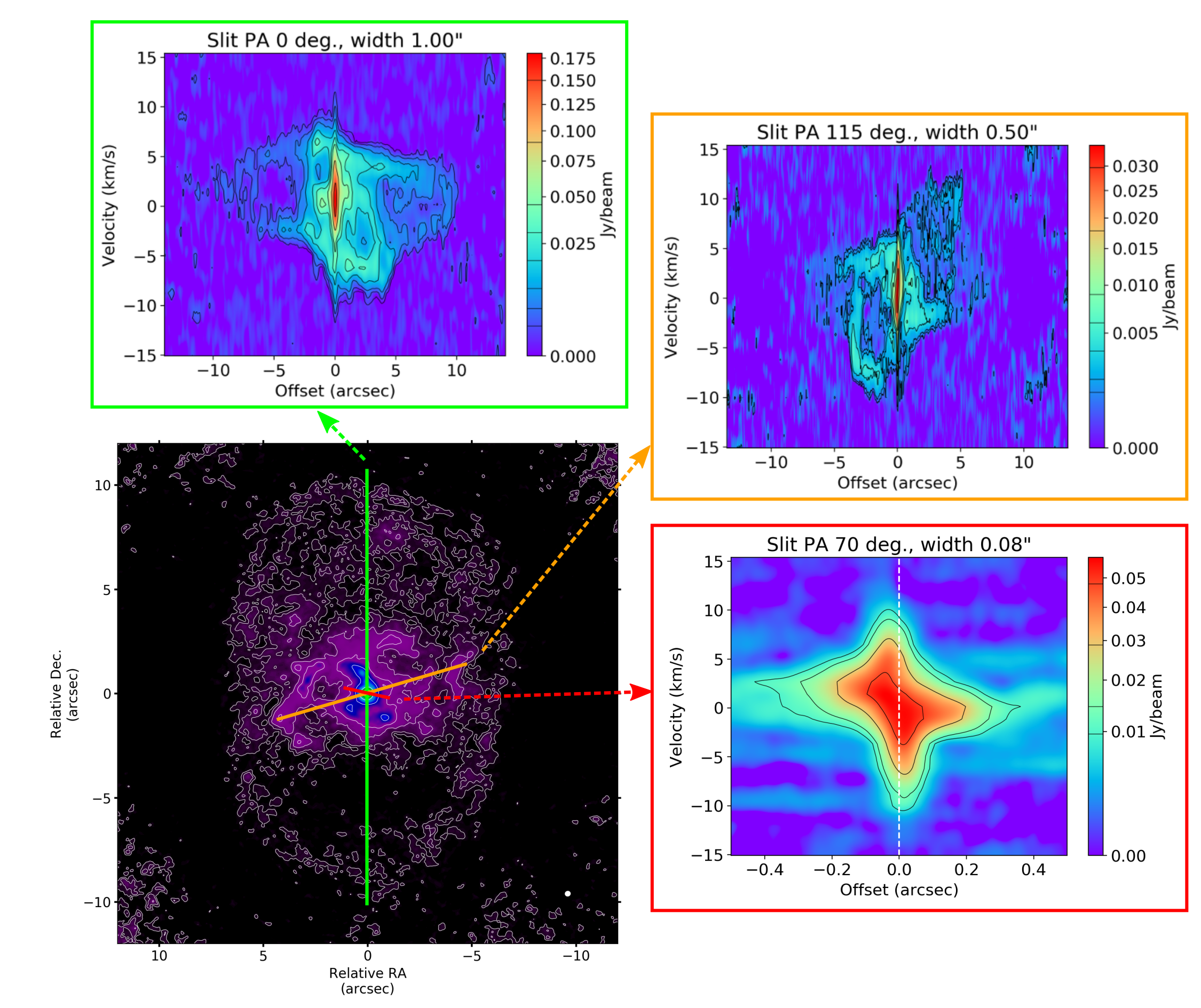}
        \caption{\emph{Bottom-left:} Central channel of the LR CO cube, with colour-coded symmetry axes obtained from the analysis in Sect. \ref{COobs}. The colour-coded encased panels represent the PV diagrams constructed along their respective colour-coded axes. The orientation and width of the slits are specified in each PV panel. Velocities are corrected to the systemic velocity of -10.1~$\kms$.}
        \label{COpvd}
    \end{figure*}

    \begin{figure}[htp!]
        \centering
        \includegraphics[width=9cm]{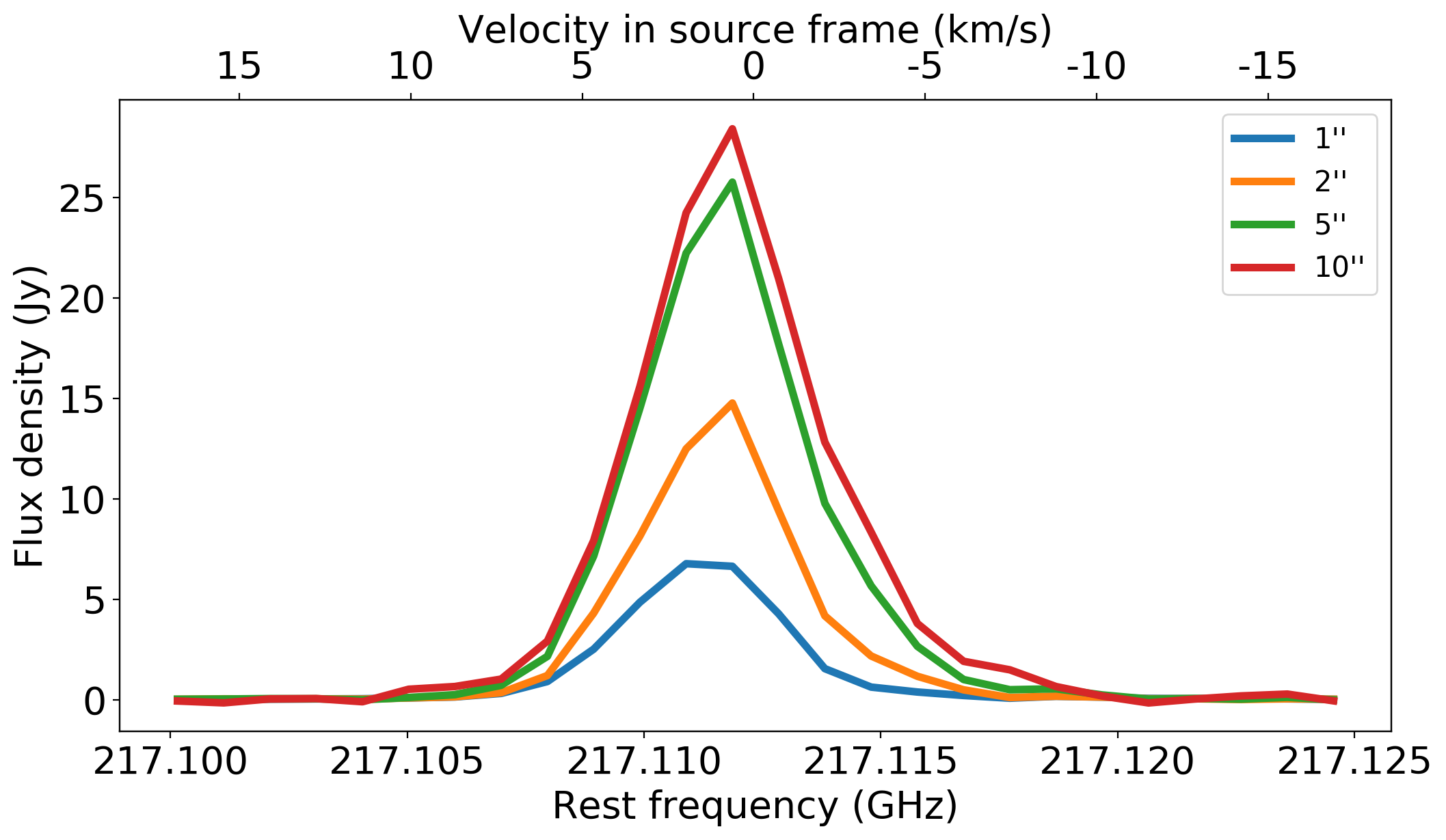}
        \caption{Line spectrum of the SiO v=0 $J=\ $5$-$4 spectral line observed in the combined dataset with different aperture diameters.}
        \label{SiOline}
    \end{figure}
        
        \subsection{CO $J=\ $2$-$1 emission} \label{COobs}
        
        The CO $J=\ $2$-$1 spectral line profile of the LR CO cube is shown in Fig. \ref{COline} for a range of different aperture sizes. These were chosen based on the angular extent of the emission (see Fig. \ref{COchan1}). It is asymmetric around the stellar velocity, best seen at the largest extraction aperture. The bulk of the emission is visually confined to velocities between $\sim$-10~$\kms$ and $\sim$14$\kms$. The line emission remains above the 3$\sigma_{\rm rms}$ noise level up to speeds of $\sim$15~$\kms$. The total integrated line flux for the largest aperture is $\sim$210~Jy$\kms$.
        
        The {\sc atomium} observations did not include the ALMA compact array (ACA) or total power, which implies that we are missing a portion of the flux at the 
largest scales. We estimate the degree of lost flux by comparing the ALMA data with single-dish SEST data \citep[Fig. 10 in][]{Winters2003}. The SEST line has an integrated flux of $\sim$330~Jy$\kms$, which implies that we are missing approximately 40\% of the total flux of the CO line. The analyses of the morphological features below focus on structures with typical length scales much smaller than the MRS of the compact configuration. We therefore do not expect this missing flux to affect the results presented in this work.
        
        We show the LR resolved emission maps of the CO line in Fig. \ref{COchan1}. The channel maps reveal a sequence of embedded structures that we describe from large to small. Between $\sim$4 and $\sim$-4~$\kms$, the channel maps are dominated by weak emission (around 3--6 $\times \sigma_{\rm rms}$) that traces a large oval feature that is oriented along the north--south axis. It has a length of $\sim$19'' and a width of $\sim$13''. To the south, a large void can be discerned, where emission drops below 3$\times \sigma_{\rm rms}$. This void has a maximal surface of $\sim$6''$\times$6'', and is confined to the south by a curved strip of faint 3$\times \sigma_{\rm rms}$ emission with a thickness of $\sim$2''. Globally, this faint emission forms the outline of an all-encompassing elliptical nebula.
        
        In the redshifted channel at 4.3~$\kms$, the extent of the southern portion of the elliptical nebula is larger than its northern counterpart, whereas in the blueshifted channel at -4.6~$\kms$ the opposite trend can be seen. Assuming that the stream lines of the wind in the elliptical nebula are predominantly radial, this implies it is slightly inclined with respect to the line of sight (with its northern part pointing towards Earth). Because of projection effects, we do not have a proper grasp on the magnitude and nature of the velocity field in the elliptical nebula at this stage, and so we cannot estimate the actual inclination angle of the nebula.
        
        The smaller scale features in the centre of the elliptical nebula can be better appraised in the HR CO channel maps shown in Fig. \ref{COchan2}. These reveal thin-walled, hollow structures in the line wings. The emission in the redshifted line wing (5 to 13~$\kms$) reveals an elongated feature with a maximal length of $\sim$5''. The size of this \emph{bubble}-like feature diminishes as speeds approach the systemic velocity. In the blueshifted line wing (-3 to -13~$\kms$), a pronounced shell of gas can be seen to the northwest of the AGB star. The shell has a thickness of $\sim$0.75'' and also forms a \emph{bubble}-like structure with a maximal diameter of $\sim$3.5''. Between 2 and -2~$\kms$ the eastern and western bubbles co-exist in the channels. 
        
        We show the \emph{moment 0} (i.e. the velocity-integrated emission pattern) and \emph{moment 1} (i.e the intensity-weighted velocity) maps of the HR CO cube in Fig. \ref{COmom0} and Fig. \ref{COmom1}, respectively. These provide a global view of the relative positions of the eastern and western bubbles. These features are separated along an axis that makes an angle of $\sim$25$^\circ$ with respect to the east--west axis. Hence, the position angle (PA) of this alignment axis is $\sim$115$^\circ$, measured from north to east.
                
        The distribution of emission in the inner 1'' of the wind is shown in Fig. \ref{COchanZ}. The brightest redshifted emission is located directly to the east of the AGB star, while its blueshifted counterpart can be found exclusively to the west. The highest velocity at which this trend can be followed is approximately $\pm$11~$\kms$, beyond which the emission is below the detection limits. Such a trend in velocity space has been observed in the inner winds of multiple AGB stars \citep{Kervella2016,Homan2018,Homan2018b}, and is typically associated with a (potentially rotating) equatorial density enhancement (EDE) of gas \citep{Homan2016}.
        
        At speeds up to $\pm$9~$\kms$, the compact redshifted (blueshifted) emission is concentrated slightly to the north (south) of the east--west axis through the position of the AGB star. The PA (measured from north to east) of this tilt can be derived by connecting the peaks of these patches at velocities around $\pm$9~$\kms$ by a line through the AGB star, which is measured to be $\sim$70$^\circ$. Under the assumption of an EDE, this would correspond to the PA of its projected major axis. In channels closest to the systemic velocity, the brightest emission is within 0.2'' north of the star, indicating that the EDE could be inclined. We discuss this further in Sect. \ref{compactdisk}. Finally, in the channels at 1.8 and 0.5~$\kms$, an unresolved \emph{hole} is seen in the emission about 0.3'' to the west of the AGB star. As discussed in Sect. \ref{SiOobs}, 
        this feature reappears in the SiO 
        emission. We further discuss this feature in Sect. \ref{void}.
        
        From the largest to the smallest spatial scales, the CO emission displays a sequence of embedded structures with apparently different alignment axes. We summarise these findings in Fig. \ref{COpvd}, where we show the position--velocity (PV) diagrams constructed along these different axes. The PV diagram along the \emph{green axis} (slit width 1'') highlights the large, elliptical, weak emission that extends $\sim$10'' to the north and south of the AGB star. The \emph{orange-coded} PV diagram (slit width of 0.5'') traces the axis along which the bubbles are separated (see Fig. \ref{COmom1}). This PV pattern emphasises the difference between the morphologies and dynamics of the eastern and western bubbles. The \emph{red-coded} PV diagram (slit width of 0.08'') probes the gas dynamics of the inner wind. Its correlated emission pattern is readily recognisable as the typical signature of gas confined to a compact EDE, as long as it is not oriented face-on, i.e. with the plane of the disc perpendicular to the line of sight \citep{Homan2016}. We further discuss the relative positioning of these features in Sect. \ref{orient}.
        
        \begin{figure*}[htp!]
                \centering
                \includegraphics[width=12.5cm]{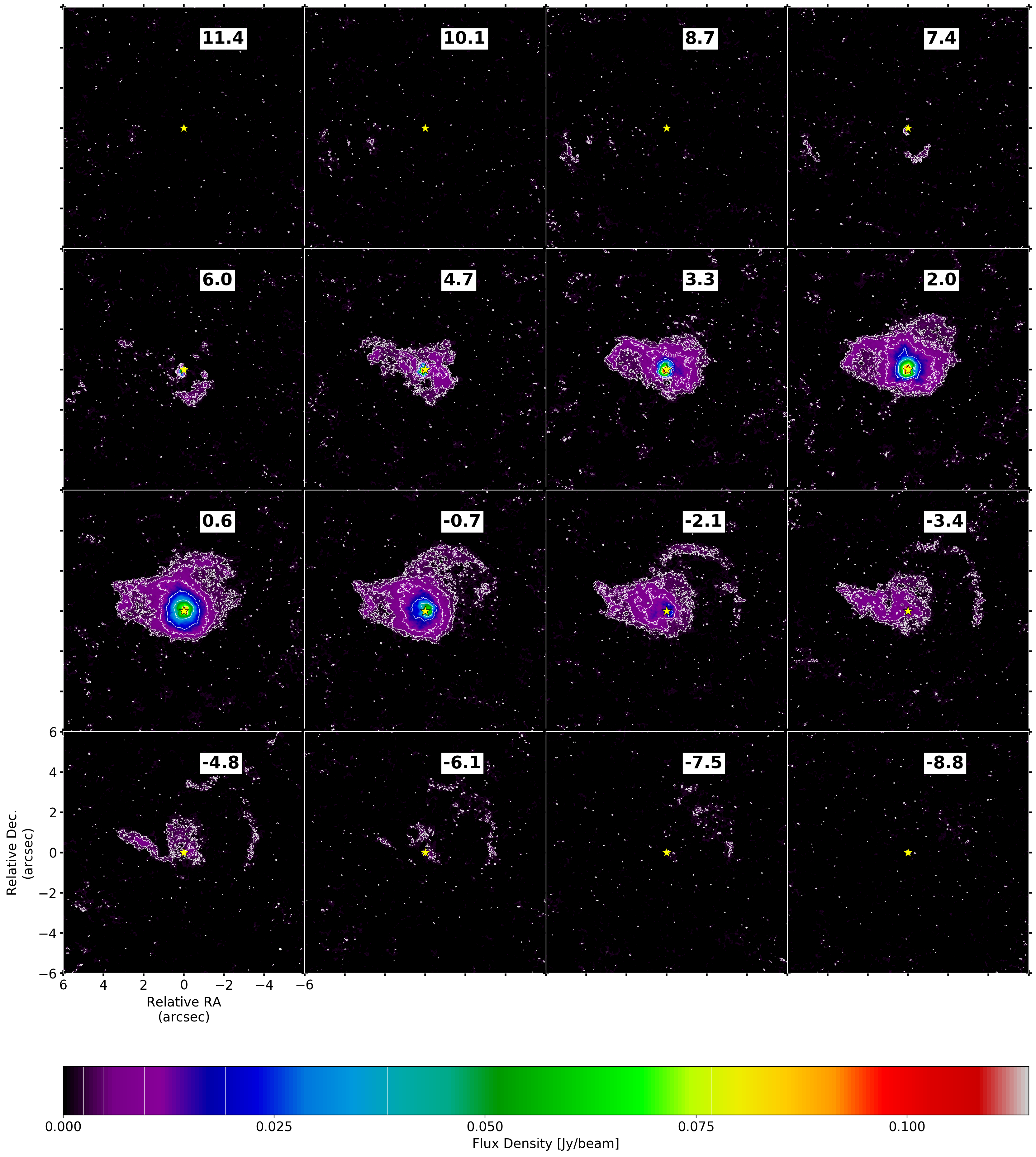}
                \caption{Channel maps showing the spatially resolved emission of the SiO v=0 $J=\ $5$-$4 line in the $\pm \sim$10~$\kms$ velocity range for the combined dataset. The velocities have been corrected for $v_{*}$=-10.1~$\kms$. Contours are drawn at 3, 6, 12, 24, 48, 96, and 192$\times \sigma_{\rm rms}$ ($\sigma_{\rm rms}$ = 0.7$\times {\rm 10}^{\rm -3}$ Jy/beam). Angular scales are indicated in the bottom left panel. The ALMA beam has a size (0.088''$\times$0.066''). The maps are centred on the continuum peak at position (0,0), and are indicated by a small yellow star.}
                \label{SiOchan}
        \end{figure*}
        
        \subsection{SiO emission}
        
        \subsubsection{SiO v=0 $J=\ $5$-$4 resolved maps} \label{SiOobs}
        
        The spectral line of the SiO emission is shown in Fig. \ref{SiOline} for different aperture sizes selected based on the angular extent of the emission (see Fig. \ref{SiOchan}). It has a triangular shape, peaks around $\sim$1~$\kms$, and possesses a broad blue wing. Very weak line wings remain above the 3$\sigma_{\rm rms}$ noise level up to velocities of -18~$\kms$ and 16~$\kms$. This is substantially larger than the velocity-extent of the CO line wings, but the bulk of the SiO emission has a velocity range of $\sim \pm$8~$\kms$ which is considerably narrower than the width of the bulk of the CO line.
        
        The spatially resolved emission maps are shown in Fig. \ref{SiOchan}. As expected from SiO \citep{Kervella2016,Homan2018b}, the bulk of the emission (greater than 24$\times \sigma_{\rm rms}$) fully encompasses the stellar position within a radius of $\sim$2'' ($\sim$300~au). In addition, some emission traces the outline of the blueshifted (western) bubble observed in the HR CO cube (see Fig. \ref{COchan2}), which manifests itself in the channel maps as a thin curved arc at velocities between $\sim$0 and $\sim$-7$\kms$. Similarly, the redshifted emission to the east of the AGB star also appears to partly outline the northern edge of the eastern bubble seen in CO (Fig. \ref{COchan2}). The easternmost edge of this bubble is also visible up to speeds of $\sim$10~$\kms$ at the 3$\sigma_{\rm rms}$ level, and, as in CO (see Fig. \ref{COchan2}), is found at an angular distance of $\sim$5'' from the AGB star. This could indicate that the SiO emission is tracing a shock, which we further elaborate on in Sect. \ref{shocks}. We also notice that an unresolved void appears in the channel at 0.6~$\kms$, about 0.3'' to the west of the AGB star (see Fig. \ref{hole}), such as was detected in CO (see Sect. \ref{COobs}). We further discuss this is Sect. \ref{void}.
        
\begin{figure*}[]
                \centering
                \includegraphics[width=8.5cm]{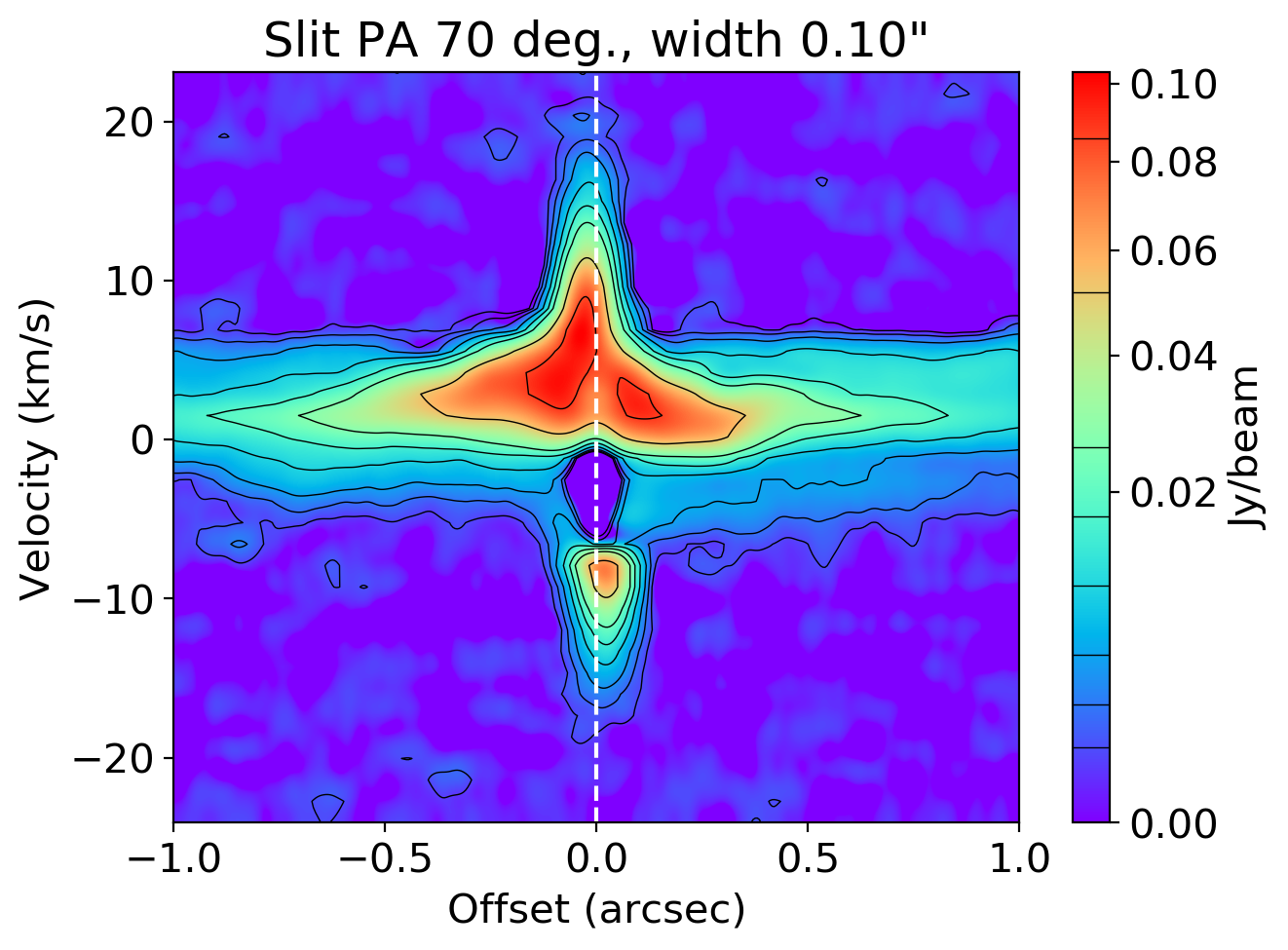}
                \includegraphics[width=8.5cm]{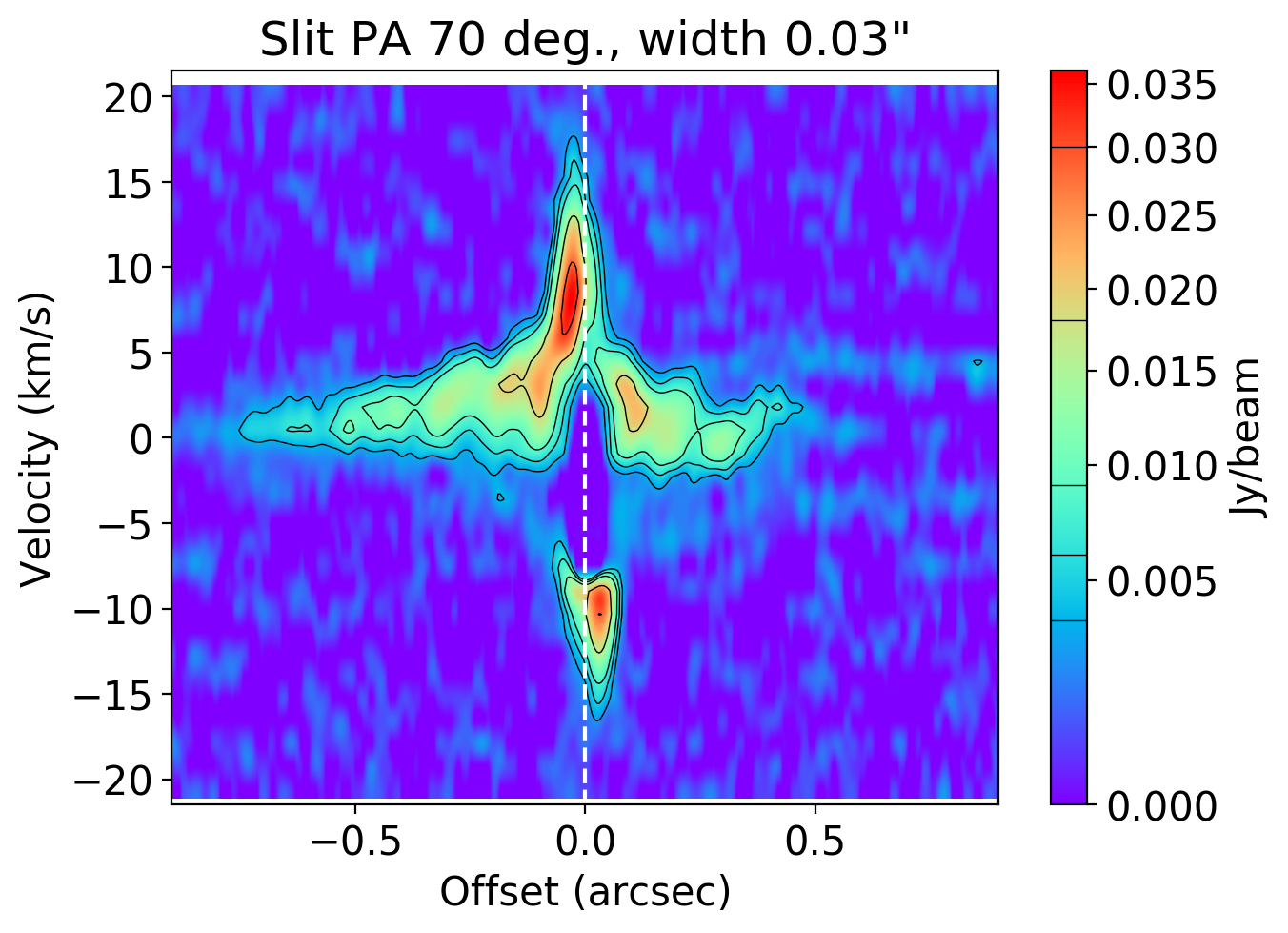}
                \caption{\emph{Left:} PV diagram of the SiO combined dataset, extracted from a slit with a width of 0.1'' and a PA of 70$^\circ$. \emph{Right:} PV diagram of the SiO extended array data, extracted from a slit with a width of 0.03'' and a PA of 70$^\circ$. The white dashed line in both panels represents the position of the AGB star.}
                \label{SiOpvd}
        \end{figure*}

        Analogous to the trend observed in the inner wind of the HR CO cube (see Sect. \ref{COobs}), compact emission is detected adjacent to the position of the AGB star. To better visualise the distribution of this emission close to the AGB star, we extracted a PV diagram of this emission for a slit width of 0.1'' along an axis with a PA of 70$^\circ$ in Fig. \ref{SiOpvd} (left panel), consistent with the PA axis determined from CO (see red axis in Fig. \ref{COpvd}). The compact emission remains detectable up to projected speeds of $\sim$18~$\kms$, explaining the broad weak wings of the spectral line. Such broad symmetrical wings in the PV diagram are typically associated with differential rotation of an EDE. This is further investigated in Sect. \ref{compactdisk}. However, the expected butterfly-shaped signature of such an EDE (see e.g. the PV diagram along the red axis in Fig. \ref{COpvd}) is not recovered in the current plot. This is mainly caused by two effects. 
        
        Firstly, around the systemic velocity, the bright, larger scale, radially expanding SiO emission surrounding the AGB star and the EDE overpowers the spatially smaller scale features responsible for the broad line wings. This causes the extended band of emission between $\sim$-7 and $\sim$7~$\kms$ in the diagram. Secondly, a large absorption feature is present in the 0 to -7~$\kms$ velocity range centred on the position of the AGB star. The latter is recovered in most data where the aperture diameter approaches the angular size of the star \citep{Decin2018}, and is a consequence of absorption along the lines of sight with impact factors that cross the stellar disc. Combined, these two effects highly perturb the expected EDE signature, such as seen in the PV diagram of CO (Fig. \ref{COpvd}). To mitigate the effect of the extended emission, we also extracted a PV diagram for the extended array data only, with a slit width of 0.03'' along the same axis, shown in the right panel of Fig. \ref{SiOpvd}. The channel maps are shown in Fig. \ref{SiOchanE} in the Appendix. The expected trend towards negative offsets is better recovered, but the absorption feature still disrupts completion of the expected PV pattern at positive offsets. 
        
        Both PV diagrams in Fig. \ref{SiOpvd} emphasise the smooth broad line wings, which are offset with respect to the AGB star (the white dashed line), and the redshifted (blueshifted) tail offset by $\sim$0.05'' from the stellar position. The offset in position of the extreme velocity emission is significant (see Sect. \ref{masers}), which means that it cannot arise from the stellar atmosphere whose  radius is 0.013''(see Sect. \ref{continuum}). This is even more apparent in the higher resolution extended-only configuration SiO isotopologue data shown in the following section, providing additional support to the hypothesis that the inner wind of R~Hya contains a differentially rotating disc, which we fully develop in Sect. \ref{compactdisk}. 
        
        The moment maps for this line of SiO do not add to the current descriptions, but for the sake of completeness these are shown in Fig. \ref{SiOmom}.
        
        \begin{figure}[]
                \centering
                \includegraphics[width=8.5cm]{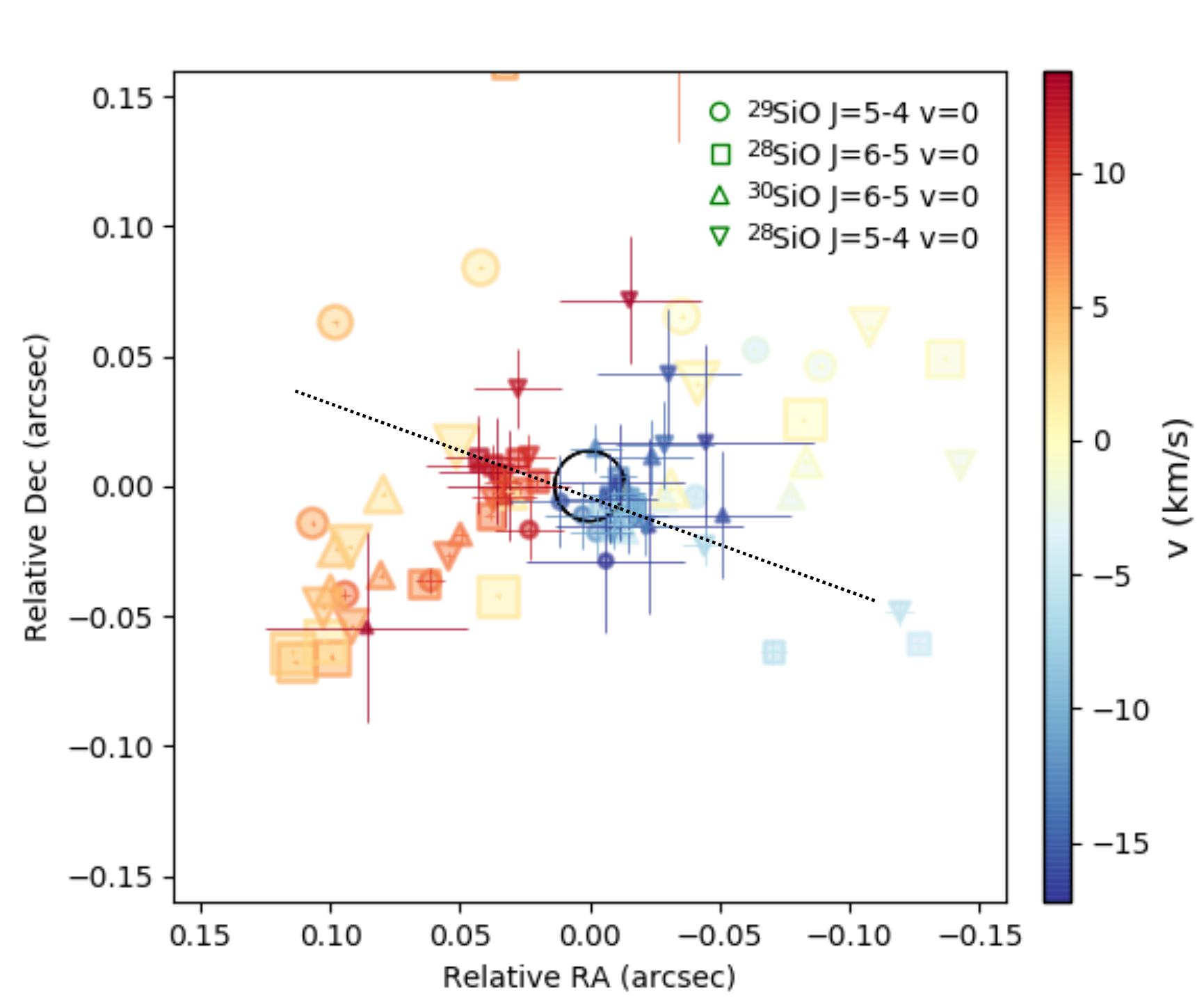}
                \caption{Position and velocity of the peak positions of the Gaussian-fitted components of the $^{\rm 28}$SiO,$^{\rm 29}$SiO $J=\ $5$-$4, and $^{\rm 28}$SiO, $^{\rm 30}$SiO $J=\ $6$-$5 ground-state vibrational emission lines of the mid-configuration dataset. The black circle in the middle represents the size of the stellar photosphere (Sect. \ref{continuum}). The marker size reflects the magnitude of the integrated flux of the Gaussian-fitted component, scaled as log$_{\rm 10}$(integrated flux). The dashed black line indicates the PA=70$^\circ$ axis.}
                \label{masers0M}
        \end{figure}
        
        \begin{figure}[]
                \centering
                \includegraphics[width=8.5cm]{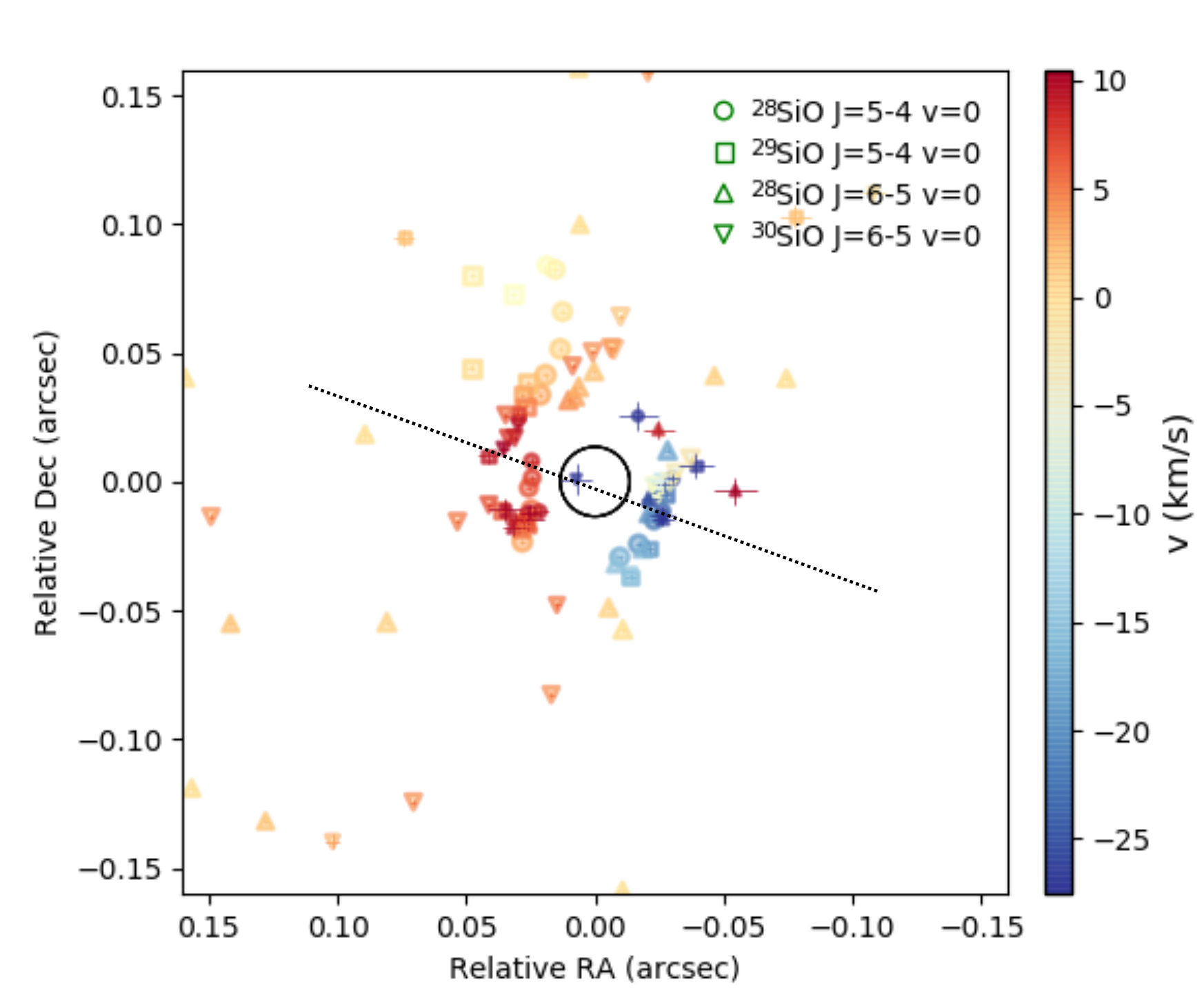}
                \caption{Same as Fig. \ref{masers0M}, but for the extended configuration dataset. We note that the velocity scale differs from Fig. \ref{masers0M}.}
                \label{masers0E}
        \end{figure}

        
        \subsubsection{Smallest scale SiO emission} \label{masers}
        
    The dynamics of the inner wind were mapped by fitting 2D Gaussian components to the peaks of the rotational lines in the ground and vibrationally excited levels of the silicon isotopologues of SiO, which is analogous to what was done in the analysis of the SiO emission of the AGB star $\pi^{\rm 1}$ Gruis \citet[][see Sect. 4.4 in {Homan2020b}]. We concentrate on the v=0 transition as the spectra suggest that it is masing only weakly if at all (Pimpanuwat et al., \emph{in prep.}), thereby simplifying their interpretation. We show the results in Figs. \ref{masers0M} and \ref{masers0E}, which were constructed using the mid and extended configuration data, respectively. These data support the trends observed in the CO and SiO data: they trace redshifted emission primarily to the east, and blueshifted emission primarily to the west of the AGB star. 
        
        In Fig. \ref{masers0M} a radial trend can be seen within a radius of $\sim$0.15'' ($\sim$25~au), with points closer to (farther from) the AGB star tracing higher (lower) speeds. In addition, the more distant data points to the southeast appear to be situated along an axis with a PA of 115$^\circ$, similar to the orange axis in Fig. \ref{COpvd}. The PA along which the inner emission is separated is 
        about $\sim$70$^\circ$, plotted as the black dashed line. The positional uncertainties of mid-configuration data is rather large. These can be mitigated by instead considering the extended configuration dataset shown in Fig. \ref{masers0E}.
        
        The extended configuration data reveal that the distribution of the innermost compact emission follows a ring-like structure, with an approximate radius of $\sim$0.04'' ($\sim$6~au). This is well beyond the stellar photosphere, which is located at 0.0135'' (see Sect. \ref{continuum}), and so the extreme velocity emission is located at $\sim$3 stellar radii. The detection of a ring implies that the emission does not originate from a structure that fully surrounds the star, such as a shell. The redshifted points are located to the east of the AGB star, and can be traced northwards, whereas the blueshifted points are located to the west, and can be followed southward. The highest speeds are traced by data points located on an axis with a PA of 70$^\circ$ to within the uncertainties consistent with the PA determined from the CO emission (see Sect. \ref{COobs}). The speed traced by the points decreases with distance away from this axis. These patterns are indicative of rotation, because the highest speeds in a radial field would be found along the north-to-south axis. In addition, the detection of speeds of such magnitude implies that the axis of rotation cannot be parallel to the line of sight. Thus, these patterns are highly consistent with gas confined to a flattened, inclined, rotating reservoir of gas, such as that supported by the other SiO and CO data (see Sects. \ref{SiOobs} and \ref{COobs}). We summarise the evidence for the presence of a rotating disc in Sect. \ref{compactdisk}.
        
        As can be seen in Fig. \ref{masers0E}, the ring is interrupted by a gap to the north and south, measured at a radius of 0.04''. The gap to the north has a width of 0.013''. Due to its small size, the significance of the gap to the north can be questioned, especially considering that the addition of one well-placed data point would negate its presence in the map. The gap to the south has an angular size of 0.03'', which is more than twice as large as the northern gap. These gaps appear along an axis that is perpendicular to the axis with a PA of 70$^\circ$, along which the highest CO velocities are separated (see red axis in Fig. \ref{COpvd}). The thermal nature of the emission (i.e. probably not masing) implies that these gaps are probably unrelated to optical or beaming effects associated with masers. 
        Considering that the SiO emission may be tracing a rotating disc, the gaps could therefore simply be the consequence of a geometrical effect. We elaborate on this discussion in Sect. \ref{compactdisk}.
        
        For the sake of completion, Fig. \ref{masers1E} also provides the position of the compact v>0 emission in relation to the compact v=0 emission. It is highly convoluted and at this stage too difficult to interpret in the context of what is presented in this manuscript.

        \section{Discussion} \label{discus}

        \subsection{Compact disc in the inner CSE} \label{compactdisk}
        
        \begin{figure}[]
                \centering
                \includegraphics[width=9cm]{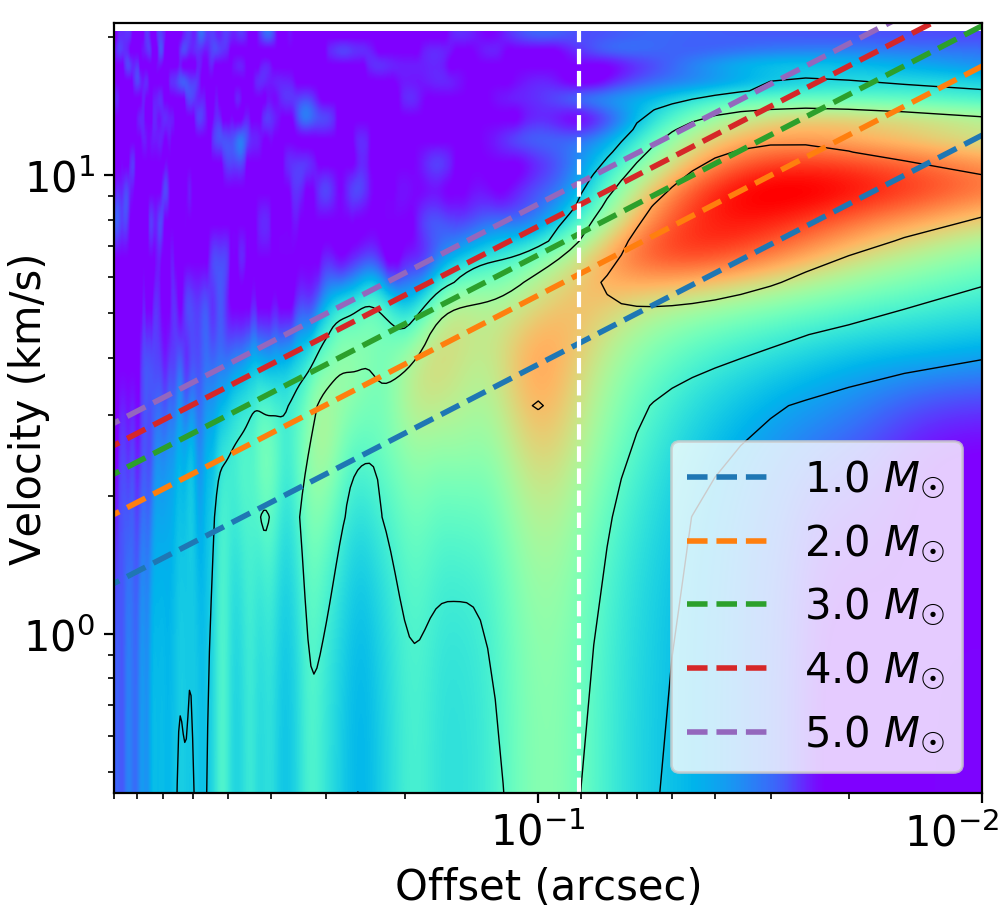}
                \caption{First quadrant of the PV diagram of the extended SiO data (Fig. \ref{SiOpvd}, right panel) in log-scale. The dotted lines represent Keplerian trends for the respectively colour-coded central masses. The vertical, white dashed line represents the position of the ring of compact SiO emission shown in Fig. \ref{masers0E}, which is assumed to be the inner rim of the disc. The offset is measured with respect to the stellar position. The contours are drawn at 3, 6, and 24 times the noise rms. At offsets larger than the inner rim position, the 3$\times \sigma_{\rm rms}$ contour lies between the 2 and 4~$\mso$ Keplerian trends, following the 3~$\mso$ trend quite closely.}
                \label{SiOpvdL}
        \end{figure}
        
    The broad line wings seen in the PV diagrams of the SiO emission (Fig. \ref{SiOpvd}) show that the attained projected speeds close to the star reach values as high as $\sim$18$\pm$1.35~$\kms$ (see Sect. \ref{SiOobs}) which significantly exceeds the terminal speed of the  wind, namely $\sim$13$\pm$1.35~$\kms$; here the uncertainty is determined by the channel width of the cube (see e.g. Figs. \ref{COchan1} and \ref{COchan2}). This terminal speed is measured from the highest velocity emission above 3$\times \sigma_{\rm rms}$, which is located outside of the inner regions of the wind, at radii greater than 1''. For dust-driven winds, dust may not be fully formed in the inner wind, and the wind is presumably still undergoing acceleration. Hence the gas dynamics probed by the SiO emission is hard to reconcile with the kinematics of a radial, slowly accelerating, dust-driven wind.
        
        The PV diagram from the inner portion of the CO maps (see red-coded panel in Fig. \ref{COpvd}) reveals the unmistakable pattern associated with a compact EDE \citep{Homan2016}. In addition, the broad line wings of SiO (see Fig. \ref{SiOpvd}) suggest that projected speeds gradually increase as one approaches the position of the AGB star. Furthermore, the compact SiO emission plots in Figs. \ref{masers0M} and \ref{masers0E} reveal that the highest projected speeds lie along an axis with a PA of $\sim$70$^\circ$, which is consistent with the PA of the projected semi-major axis of the EDE signal extracted from the CO PV diagram (see red-coded panel in Fig. \ref{COpvd}). Finally, the compact SiO emission traces the edges of a ring-like pattern, along which speeds gradually diminish as one approaches the perpendicular projected axis. Taking all these elements into consideration, we propose that the compact emission in the inner wind traces a differentially rotating disc.
        
        The compact emission traced by the data shown in Fig. \ref{masers0E} is thought to trace the most energetic regions in the inner wind, implying that they are tracing a highly dynamical, circular phenomenon inside the disc. Being the most dynamically active zone in a disc, we believe that this ring could be tracing the disc's inner rim.
        
        The brightest emission in the EDE curves over the northern side of the position of the AGB star (see Fig. \ref{COchanZ}). Presumably originating from the dense and warm interior of the disc, this implies that the disc is somewhat inclined, with its southern part pointing towards Earth. The contrast between the emission to the north and south of the AGB star (around e.g. 0.5~$\kms$ in Fig. \ref{COchanZ}) indicates that the inclination angle is probably not large. This also indicates that the disc could be flared, because at lower inclination angles (with respect to the line-of-sight) the southern part of the inner rim remains obscured by the cooler gas in the disc at larger radii. This was for example also seen in the disc around L$_{\rm 2}$ Puppis \citep{Homan2017}. This type of obstruction could also explain the southern gap in Fig. \ref{masers0E}, and its position along an axis perpendicular to the axis with a PA of 70$^\circ$ (see Fig. \ref{COpvd}). 
        
        Rotating discs are thought to be generally found around binary systems becsuse these systems have access to the significant angular momentum reserves stored in the orbit of the  companion \citep{Homan2017,Chen2017}. The known companion of R~Hya, located at an angular separation of 21'' from the AGB star \citep{Mason2001}, affects the nebular gas on timescales that are much longer than the dynamical timescale of the inner wind. Hence, this companion is unable to affect the gas dynamics of the inner wind in any significant way. We therefore propose that R~Hya possesses another, much closer companion. 
        
        Assuming a first-order inclination of $\sim$30$^\circ$ (see also Sect. \ref{orient}), we can estimate the central mass of the companion using Keplerian dynamics. The ring of compact SiO emission seen in Fig. \ref{masers0E} has a physical radius of $\sim$6~au, which we believe to be tracing the inner rim of the disc. For projected speeds of $\pm$18~$\kms$ to be measured at this location, the central mass of the system would have to be $\sim$2.5~$\mso$. The mass estimate of $\sim$1.35~$\mso$ by \citet{Decin2020} is based on isotopic ratios of oxygen, which carefully trace the properties of the AGB star. As such, we predict that the mass of the closeby companion is close to 1~$\mso$. This is also supported by Fig. \ref{SiOpvdL}, where we plotted the first quadrant of the PV diagram of the extended SiO data (Fig. \ref{SiOpvd}, right panel) in log-scale. The white dashed line represents the position of the inner rim, and any signal closer to the star is not resolved and cannot be trusted. Overplotted are the expected Keplerian trends for five different central masses. The trend that best matches the highest contour (3$\times \sigma_{\rm rms}$) corresponds to a central mass of between 2 and 4 solar masses, yielding a lower limit on the companion mass of 0.65~$\mso$.
        
        Though we assume that the gas at the inner rim is undergoing Keplerian rotation, the gas farther out in the disc may be, in addition to rotating, also expanding and/or ablating, but the degeneracy between the orientation of the disc on the one hand, and tangential versus radial velocity on the other, means that these cannot be fully disentangled in the current dataset. To get a better grasp on the dynamics governing the disc, higher resolution images need to be acquired, as was done for L$_{\rm 2}$ Puppis \citep{Kervella2016,Homan2017,Haworth2018}. However, the details of the gas dynamics within the disc will not affect our current companion mass estimate.

    Finally, we want to focus the attention on an alternative scenario, which has been proposed as a different way to interpret these patterns. As suggested by \citet{TuanAnh2019} for EP~Aquarii, and by \citet{Do2020} for R~Doradus, the high-velocity features detected in the inner wind of these AGB stars may also be tracing short jet-like streams of gas that possess rather narrow opening angles. Provided these jets accelerate rapidly within the first few tens of astronomical units (au) and then decelerate gradually over the next hundred or so au, this explanation can serve as a valid alternative to the disc hypothesis. This invokes the question as to how these jets are launched, shifting the focus back towards the presence of an actively accreting companion \citep{Blandford1982,Rekowski2000}. As such, it seems that the presence of a companion is probably indispensable in explaining the enhanced dynamics that are detected in the inner winds of certain oxygen-rich AGB stars.
        
        \subsection{Bubbles: Energetic shock fronts} \label{shocks}
        
        The edge of the westward-pointing bubble can be seen both in CO and notably in SiO (see Figs. \ref{COchan2} and \ref{SiOchan}), up to distances of $\sim$4'', or $\sim$660~au away from the star. 
    SiO emission is typically confined to the inner portion of the wind \citep[e.g.][]{Homan2018b,Decin2018}. In addition, due to its large electric dipole moment the excitation of the SiO molecule is primarily sensitive to the stellar radiation field, while only poorly tracing the gas density. Therefore, its detection in such isolated strips, located so far away from the centrally condensed bulk of the SiO emission surrounding the AGB star, is curious.
        
        The SiO emission traces a shell thickness of only $\sim$0.3'', that is, a few beam widths, at the 3$\times \sigma_{\rm rms}$ level (see Fig. \ref{SiOchan}). Such a thin shell is indicative of a shock wave. Unfortunately, projection effects impede the estimation of the 3D speed and orientation of the bubble with respect to the AGB star. Nevertheless, some estimates can still be made. We use the Newton-Laplace equation and ideal gas law to estimate the sound speed of the medium, and to estimate whether the bubble is supersonic. By assuming a simplistic AGB radial temperature profile starting at the stellar surface of $T \sim r^{-0.5}$ \citep{DeBeck2010}, we calculate the temperature of the unperturbed AGB wind at $\sim$4'' to be $\sim$125 K. This gives a local sound speed of $\sim$0.85~$\kms$.
        The maximal projected speed in the bubble is $\sim$13~$\kms$ (see Fig. \ref{COchan2}), and so the feature is travelling through the local medium at a speed of at least Mach 15, which is well within the strong shock regime.

        SiO is known to be an excellent tracer of shocks \citep{Schilke1997}. In dusty outflow regions, where SiO is substantially depleted, shocks tend to substantially increase the sputtering rate of Si-bearing matter from the dust particles \citep{MartinPintado1992,JimenezSerra2005,JimenezSerra2008,JimenezSerra2009,Kelly2017}. In addition, shocks may also dissociate some of the more complex Si-bearing molecules in the affected zones \citep{Boulangier2019}. In addition, dense post-shock gas has been shown to undergo efficient cooling \citep{Gobrecht2016}. As a consequence, the conditions for post-shock chemistry are suitable for the efficient production of the most stable molecular species from the most abundant building blocks in the post-shock debris field, C, O, N and Si \citep{Karakas2016}. These molecules would be CO, SiO, and N$_{\rm 2}$. Such a chemical `reset' by the shock \citep{Schilke1997} could explain the detection of SiO and CO within the bubble's shell outline.
        
        The eastward bubble observed in the HR CO channel maps is primarily visible in the redshifted channels, and the westward bubble is most
visible in the blueshifted channels. Assuming that they are dynamically dominated by radial stream lines, this implies that the eastern bubble is pointing away from Earth, while the western bubble is pointing towards Earth. Though morphologically quite different, they appear to be approximately aligned in polar opposition along an axis with a PA of $\sim$115$^\circ$ (see Sect. \ref{COobs}, Figs. \ref{COmom1} and \ref{COpvd}). Further, they possess comparable projected speeds, are comparable in size, and both appear in the CO and SiO maps. Together, these findings suggest that the bubbles could be generated by a singular energetic event \citep[e.g. a TP, see][]{Zijlstra2002} that launched these shock waves, which, due to highly anisotropic local conditions, have been moulded into two morphologically different shock fronts. 
        
        Assuming that the shock front is propagating at a speed comparable to the measured gas speeds, then the post-shock gas would be residing on the inside of the shock front. As such, we can use the projected velocities and size of the shock front to put an upper limit on the age of the bubbles of $\sim$300 years. Interestingly, this estimate is within the same order of magnitude as the narrative proposed by \citet{Hashimoto1998} and \citet{Zijlstra2002}, and later supported by \citet{Decin2008}, that an acute event, triggered a few centuries ago, resulted in a momentary jump in gas and dust mass-loss rate. These latter authors suggest that this event could have been a TP. 
        
        
        
        \subsection{Compact emission void} \label{void}
        
        \begin{figure*}[htp!]
                \centering
                \includegraphics[width=8cm]{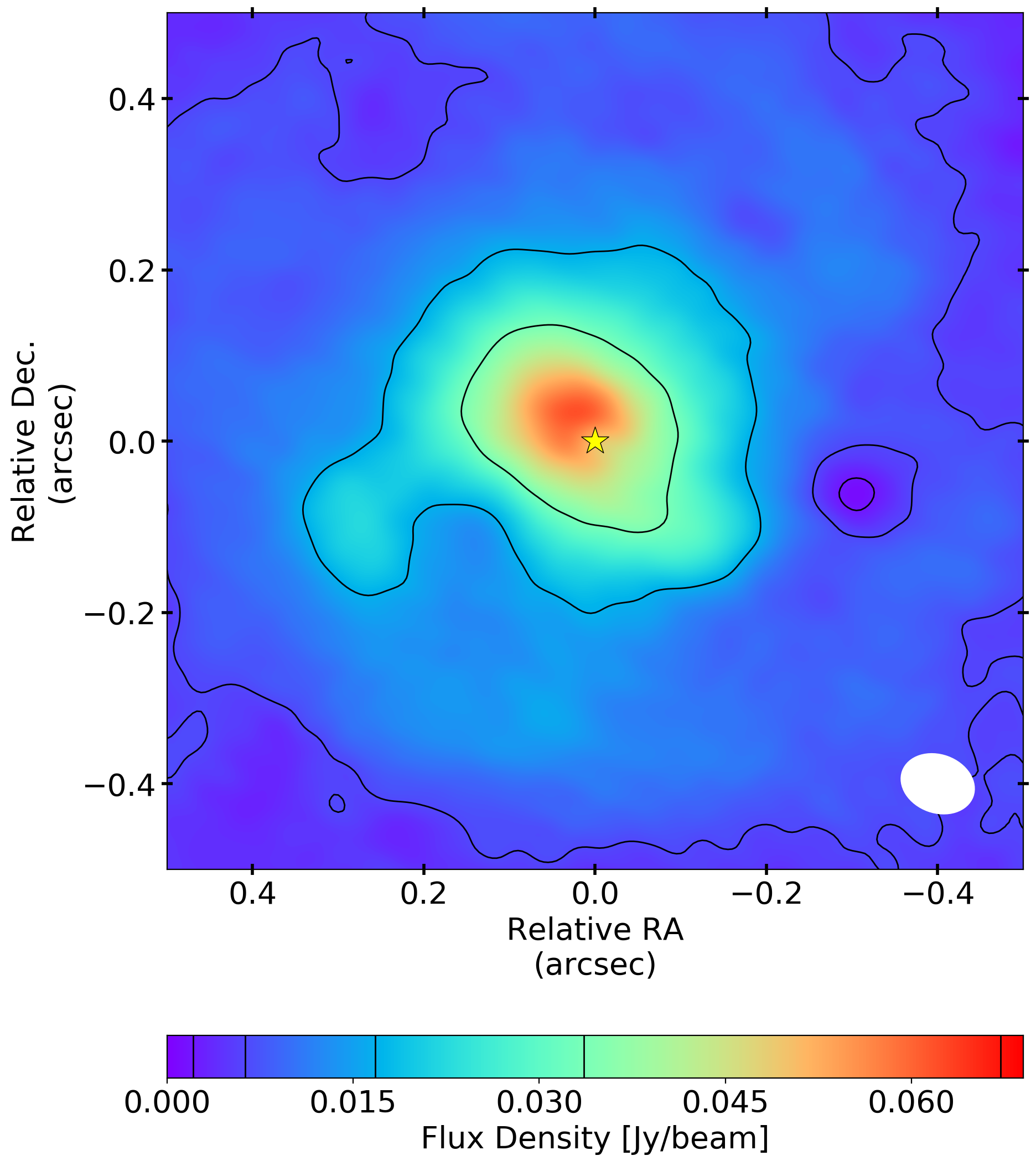}
                \includegraphics[width=8cm]{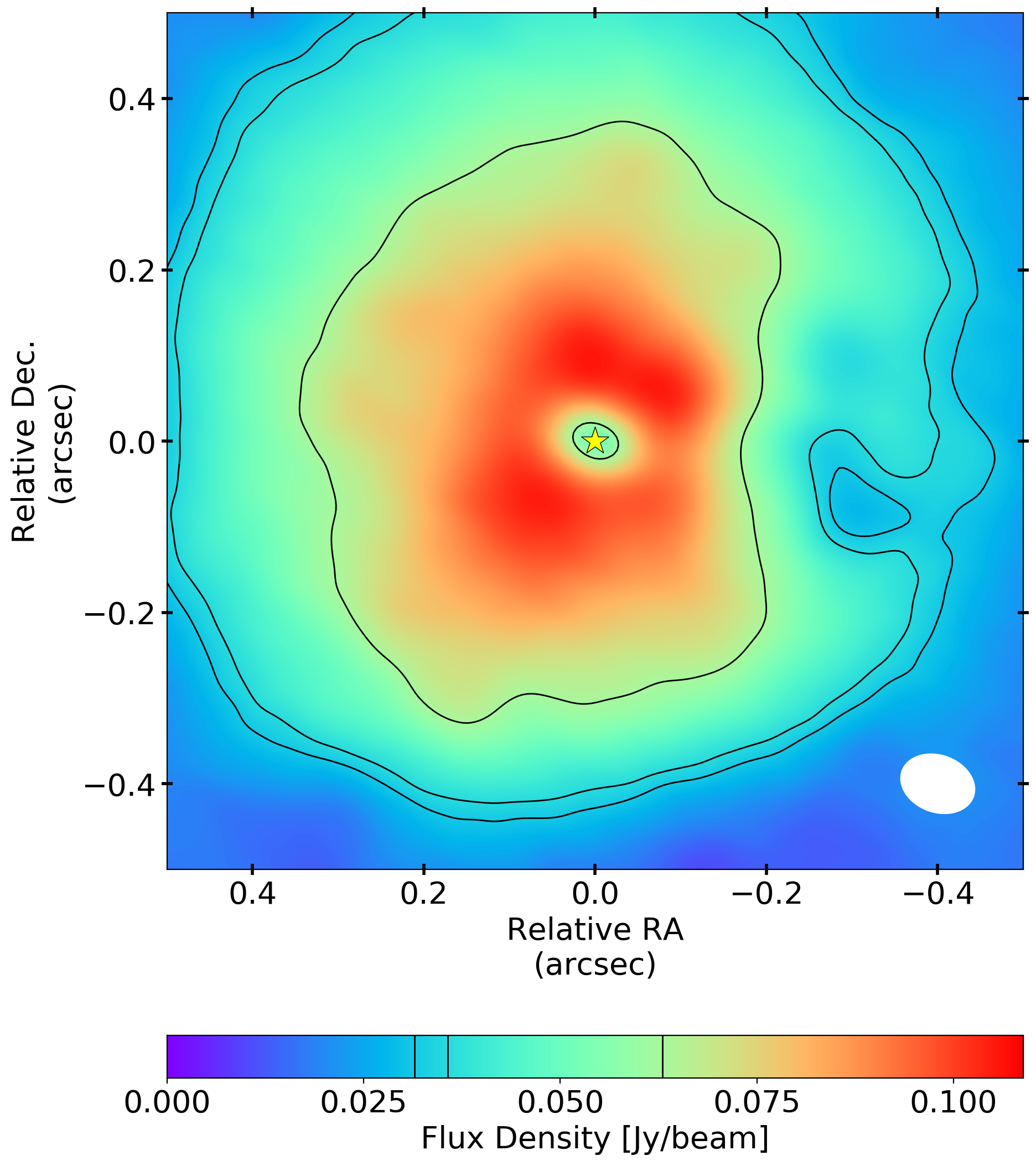}
                \caption{Channel maps where an unresolved emission void is detected about 0.3'' to the west of the AGB star (indicated by the yellow star). \emph{Left}: CO channel at 0.5~$\kms$, contour levels [3,9,24,48,96]$\times \sigma_{\rm rms}$. \emph{Right}: SiO channel at 0.6~$\kms$, contour levels [46,51,90]$\times \sigma_{\rm rms}$. The contour levels have been chosen to highlight the feature.}
                \label{hole}
        \end{figure*}

    In Fig. \ref{hole} we show how the emission void manifests itself in the channel maps of CO and SiO. The void is $\sim$0.3'' offset from the star and is therefore not caused by absorption seen towards the star. Accounting for its position in the plane of the sky, and its speed with respect to the systemic velocity, the void could be located at the edge of the rotating disc. However, the \emph{void} only appears in the redshifted portion of the spectral line, whereas the western portion of the disc is blueshifted with respect to the systemic velocity. Hence, if the void were orbiting the star, it would be doing so in the opposite direction of the disc’s rotation. We therefore conclude that the void is probably not directly related to the disc; instead the void could either be an exceptionally low-density region, or a dense clump of cool gas that resides at the outer edge of the rotating disc, slowly drifting outwards at a very moderate speed. Its projected velocity implies that it should be moving away from Earth, and is probably located behind a plane through the position of the AGB star and perpendicular to the line of sight.
    
    Alternatively, the void in R~Hya is reminiscent of a similar void in the inner wind of the AGB star EP~Aquarii \citep{Homan2018b}, which was later interpreted as a potential photodissociation region generated by the radiation field of a white dwarf \citep[WD; ][]{Homan2020}. Considering that the void in R~Hya consistently appears in approximately the same channels and at the same place in different molecules observed in the {\sc atomium} survey, the presence of a photodissociating companion (i.e. either a cool WD or hot main sequence star) cannot be immediately excluded. Depending on its properties, such a companion could easily remain hidden in spectra or SEDs, as was the case for the WD companion of EP Aqr \citep{Homan2020}, or the companion of $\pi^{\rm 1}$ Gruis \citep{Homan2020b}. However, it must be noted  that it is unlikely that R~Hya possesses two companions of significant mass within a radius of 50~au, because this would make the system highly unstable, implying that it would probably not have survived up to the AGB phase. We defer a more detailed investigation into the true nature of this void, taking into account other molecules, to a follow-up paper.

        \subsection{Relative orientation of primary wind features} \label{orient}
        
        \begin{figure}[]
                \centering
                \includegraphics[width=7cm]{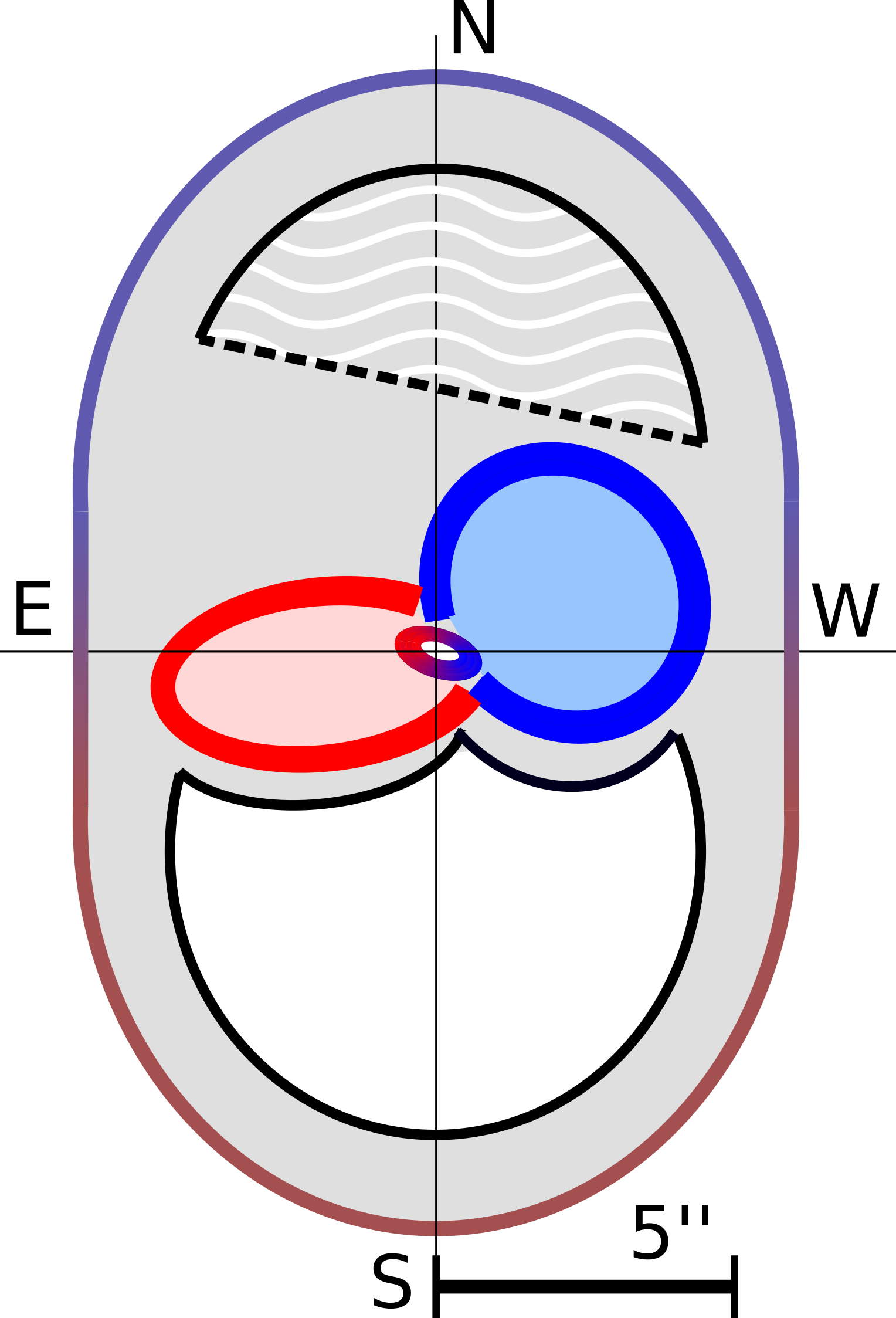}
                \caption{Schematic diagram of the main morphological features of the R~Hya nebula. All features are colour-coded to match their respective Doppler shifts. The elliptical nebula (in grey, typical size $\sim$1500~au) encompasses the eastern and western bubbles (typical size $\sim$800~au), which fully surround the compact EDE in the centre of the CSE (typical size $\sim$80~au).}
                \label{diagram}
        \end{figure}
        
        Owing to uncertainties in the velocity field, most depth information is missing.
        We therefore resort to a two-dimensional representation, and only discuss the orientation of the principal features in the R~Hya nebula with respect to each other in the plane of the sky. At subarcsecond spatial scales, the data strongly suggest the presence of an inclined rotating disc in the inner wind (see Sect. \ref{compactdisk}), shown in the centre of the schematic diagram in Fig. \ref{diagram}. The disc has an outer edge in CO of $\sim$0.4'', and an inner rim located at $\sim$0.04''. Its major axis has a PA of $\sim$70$^\circ$ in the plane of the sky and is likely moderately inclined (i.e. it is neither nearly face-on, nor nearly edge-on). At the scale of a few arcseconds, the arcs seen in CO and SiO (see Sect. \ref{COobs}) trace bubble-like shock waves that propagate through the medium in the vicinity of the AGB star. These are distinctly spectrally separated, and are colour-coded accordingly in the diagram. The \emph{blueshifted bubble} to the west is significantly rounder than the more elongated \emph{redshifted bubble} to the east, but both bubbles have very similar horizontal spatial extents, and maximal projected speeds. Finally, at the length scale of $\sim$10 arcseconds, the elliptical nebula ---represented by the encompassing {light-grey} feature in Fig. \ref{diagram}, which is confined in velocity space to the first few km s$^{-1}$ with respect to the systemic velocity--- dominates the data. The southern portion of the elliptical nebula is hollow, whereas this is less obvious in the northern segment.
        
        Hydrodynamical models of equatorially confined winds show that elliptical and bipolar morphologies can be rather easily produced \citep{Icke1988} by invoking a sudden increase in the energy injected by the AGB star. The particular shape of the resulting shock wave is sensitive to the orientation and equator-to-pole density contrast of the EDE. These models all predict that the primary symmetry axis of the resulting nebula is perpendicular to the equatorial plane of the EDE. Accounting for projection effects, it is possible that the major axis of the disc and the axis along which the bubbles are separated (see dashed line in Fig. \ref{COmom1}) are indeed approximately perpendicular to each other, provided that the inclination angle of the disc is approximately 35$^\circ$ away from edge-on. In this case, an acute energetic event, such as a TP that occurred a few centuries ago \citep{Zijlstra2002}, could explain the launching of the shocks along the orange axis in Fig. \ref{COpvd}. The inclination of the disc with respect to the line of sight can then also explain the position of the blueshifted and redshifted shock-fronts in velocity space.   
        
        This scenario cannot explain the orientation of the elliptical nebula with respect to these features. Indeed, there is currently no simple explanation for its shape and orientation that is immediately consistent with the morpho-kinematical picture of the inner wind presented in this paper. In addition, the low projected speeds of the elliptical nebula imply that it is probably propagating primarily in the direction perpendicular to the line of sight, making it impossible to estimate the actual speed of the gas, and consequently its kinematical age.
        The development of a comprehensive morphological model of the nebula is deferred until additional analyses of the {\sc atomium} data have yielded tighter constraints on the nature of the object.
        
        Finally, we want to shift the attention to the fact that the previous one-dimensional interpretations of the mass-loss history of R~Hya \citep{Hashimoto1998,Zijlstra2002,Decin2008,DeBeck2010,Schoier2013} probably need to be revised in light of what is now known about the complex structure and high velocities of the wind. For example, the apparent decline in mass-loss rate in the literature could be due to the pronounced density contrast between the bubbles and their walls, meaning that the assumptions about optical depth gradients and relative excitation of different CO transitions were not entirely applicable. Investigating the extent of these uncertainties is beyond the scope of this paper, but they are of absolute importance to better understand the mass-loss history of this target.        

        \section{Summary} \label{summ}
        
        The circumstellar environment of the M-type AGB star R~Hydrae was observed with ALMA as part of the {\sc Atomium} large programme. The target was observed with three different antenna configurations, covering spatial scales from $\sim$5 to $\sim$1500~au, allowing for an in-depth morphological study of the stellar wind. To this end, we used the continuum emission, as well as resolved maps of CO v=0 $J=\ $2$-$1 and SiO v=0 $J=\ $5$-$4 molecular line emission. The continuum is centrally condensed, showing a slight elongation in the direction of the ALMA beam. The total flux density within the 3$\times \sigma_{\rm rms}$ contour is $\sim$65~mJy. All but the shortest baselines of the continuum visibilities are very well fitted by a uniform disc with a diameter of 0.027'' (4.5~au) and a flux density of 56~mJy, which we assume to be emission from the AGB star. A small continuum excess of $\sim$10~mJy of larger scale flux, probably caused by nearby warm dust, is also present in the data. The CO emission reveals a highly complex arrangement of distinct morphological features on different length scales. At length scales of $\sim$1600~au, the CO channel maps are dominated by an elliptical feature,  which is elongated along the north--south axis. Embedded within this feature the CO emission reveals two hollow bubbles with a characteristic size of a few hundred au, one blueshifted to the west, and one redshifted to the east. These features are also recovered in the SiO channel maps, indicating that they could be energetic shock waves propagating through the nearby circumstellar medium. Estimating the dynamical age of the features to a few hundred years places them in the same approximate age range as the predicted time-interval since the star last underwent a thermal pulse. At length scales of $\sim$50~au, the CO emission surrounding the AGB star follows a typical butterfly pattern that is readily associated with the presence of an equatorial density enhancement. Analysis of the resolved SiO emission in the same region reveals that the gas in this region reached speeds of up to $\pm$18~$\kms$, which significantly exceeds the maximal wind speed measured at larger scales. This suggests that the inner wind of R~Hya could be dominated by the presence of a rotating disc, and therefore likely also harbours a close-by companion. Investigation of the compact emission of the $^{\rm 28}$SiO,$^{\rm 29}$SiO $J=\ $5$-$4, and $^{\rm 28}$SiO, $^{\rm 30}$SiO $J=\ $6$-$5 ground-state vibrational emission of the extended dataset confirms the presence of a ring-like feature with a radius of $\sim$6~au. Assuming that the highest speeds measured in the winds of the SiO v=0 $J=\ $5$-$4 molecular maps trace the inner rim of the disc, we tentatively estimated a lower limit on mass of the companion to be $\sim$0.65~$\mso$.
        
        \begin{acknowledgements}  
                
                We would like to thank the Italian ARC Node, and in particular R. Laing for providing the scripts that were used to analyse the continuum.
                
                This paper makes use of uses the following ALMA data: ADS/JAO.ALMA\#2018.1.00659.L, ‘ATOMIUM: ALMA tracing the origins of molecules forming dust in oxygen-rich M-type stars’. ALMA is a partnership of ESO (representing its member states), NSF (USA), and NINS (Japan), together with NRC (Canada),  NSC and ASIAA (Taiwan), and KASI (Republic of Korea), in cooperation with the Republic of Chile. The Joint ALMA Observatory is operated by ESO, AUI/NR.A.O, and NAOJ.
                
                W.H acknowledges support from the Fonds de la Recherche Scientifique (FNRS) through grant 40000307. 
                
                S.H.J.W., L.D, M.M., D.G. acknowledge support from the ERC consolidator grant 646758 AEROSOL. 
                
                TD acknowledges support from the Research Foundation Flanders (FWO) through grant 12N9920N. 
                
                TJM is grateful to the STFC for support via grant number ST/P000312/1.
                
                JMCP acknowledges funding from the UK STFC (grant number ST/T000287/1)           
                MVdS acknowledges support from the Research Foundation Flanders (FWO) through grant 12X6419N.
                
                S.E. acknowledges funding from the UK Science and Technology Facilities Council (STFC) as part of the consolidated grant ST/P000649/1 to the Jodrell Bank Centre for Astrophysics at the      University of Manchester.
                
                C.A.G. acknowledges support from NSF grant AST-1615847.
                
                We acknowledge financial support from “Programme National de Physique Stellaire” (PNPS) of CNRS/INSU, France.
                
                We used the SIMBAD and VIZIER databases at the CDS, Strasbourg (France)\footnote{Available at \url{http://cdsweb.u-strasbg.fr/}}, and NASA's Astrophysics Data System Bibliographic Services. This research made use of IPython \citep{PER-GRA:2007}, Numpy \citep{5725236}, Matplotlib \citep{Hunter:2007}, SciPy \citep{2020SciPy-NMeth}, and Astropy\footnote{Available at \url{http://www.astropy.org/}}, a community-developed core Python package for Astronomy \citep{2013A&A...558A..33A}.

        \end{acknowledgements}

        \bibliographystyle{aa}
        \bibliography{wardhoman_biblio}

        \IfFileExists{wardhoman_biblio.bbl}{}
        {\typeout{}
                \typeout{******************************************}
                \typeout{** Please run "bibtex \jobname" to obtain}
                \typeout{** the bibliography and then re-run LaTeX}
                \typeout{** twice to fix the references!}
                \typeout{******************************************}
                \typeout{}
        }
        
        \clearpage
        \begin{appendix}
                
    \section{Additional Figures}
    
    \begin{figure*}[]
        \centering
        \includegraphics[width=7.3cm]{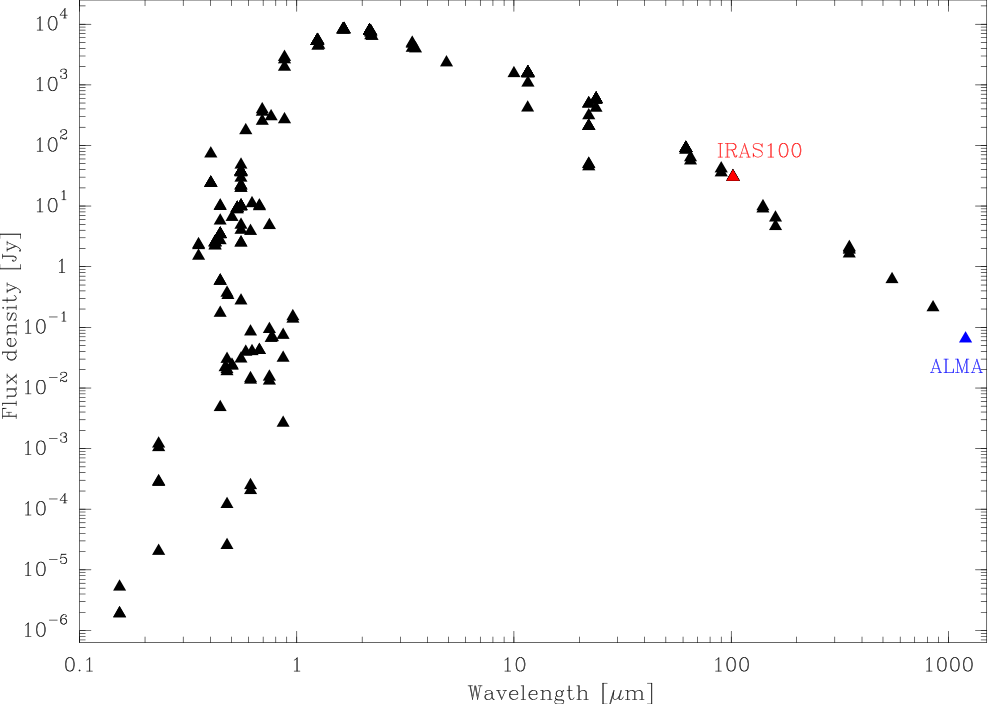}
        \includegraphics[width=7cm]{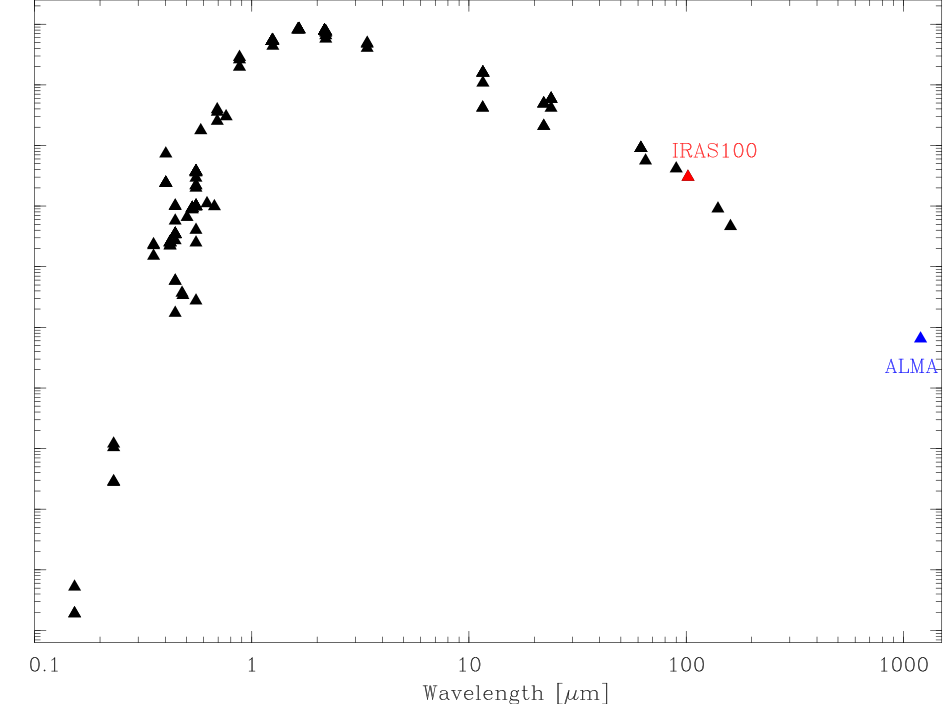}
        \caption{Spectral energy distribution for R~Hya using the continuum measurements obtained from Vizier (http://vizier.u- strasbg.fr/vizier/sed/) for a radius of 30'' (\emph{left panel}), and 1'' (\emph{right panel}), centred on the source. The flux density of the IRAS measurement at 100 microns \citep{Hashimoto1998} and of our ALMA observation at 250 GHz are shown in red and blue, respectively.}
        \label{SED}
    \end{figure*}

    \begin{figure}[]
        \centering
        \includegraphics[width=8cm]{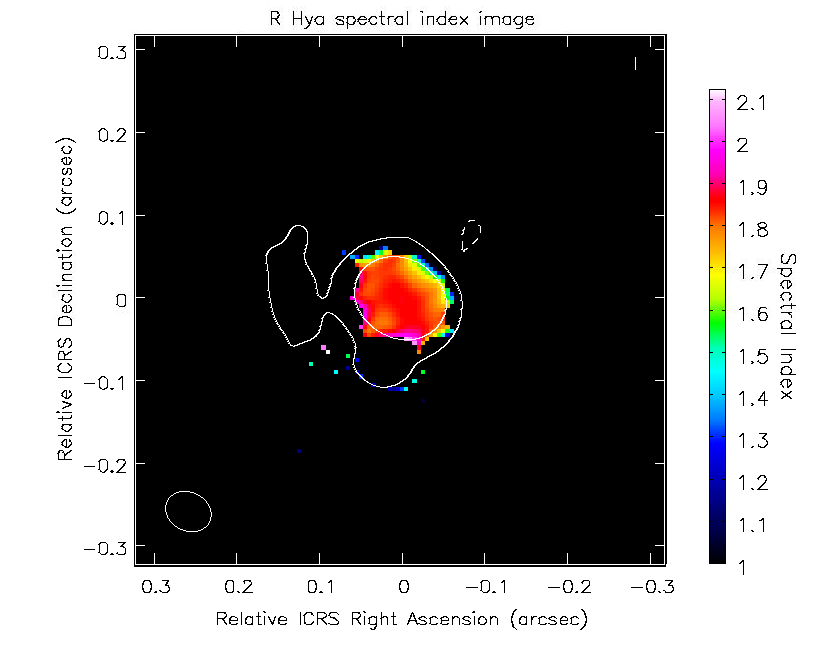}
        \caption{Spectral index between 211-275~GHz, blanked where errors are >0.2.  Contours at 6 and 96 $\times \sigma_{\rm rms}$, the inner contour corresponding approximately to a 56-mas diameter disc. This is in the typical spectral index range for AGB stars in general in the mm regime \citep{Vlemmings2019}.}
        \label{specIn}
    \end{figure}

        \begin{figure*}[]
                \centering
                \includegraphics[width=8.5cm]{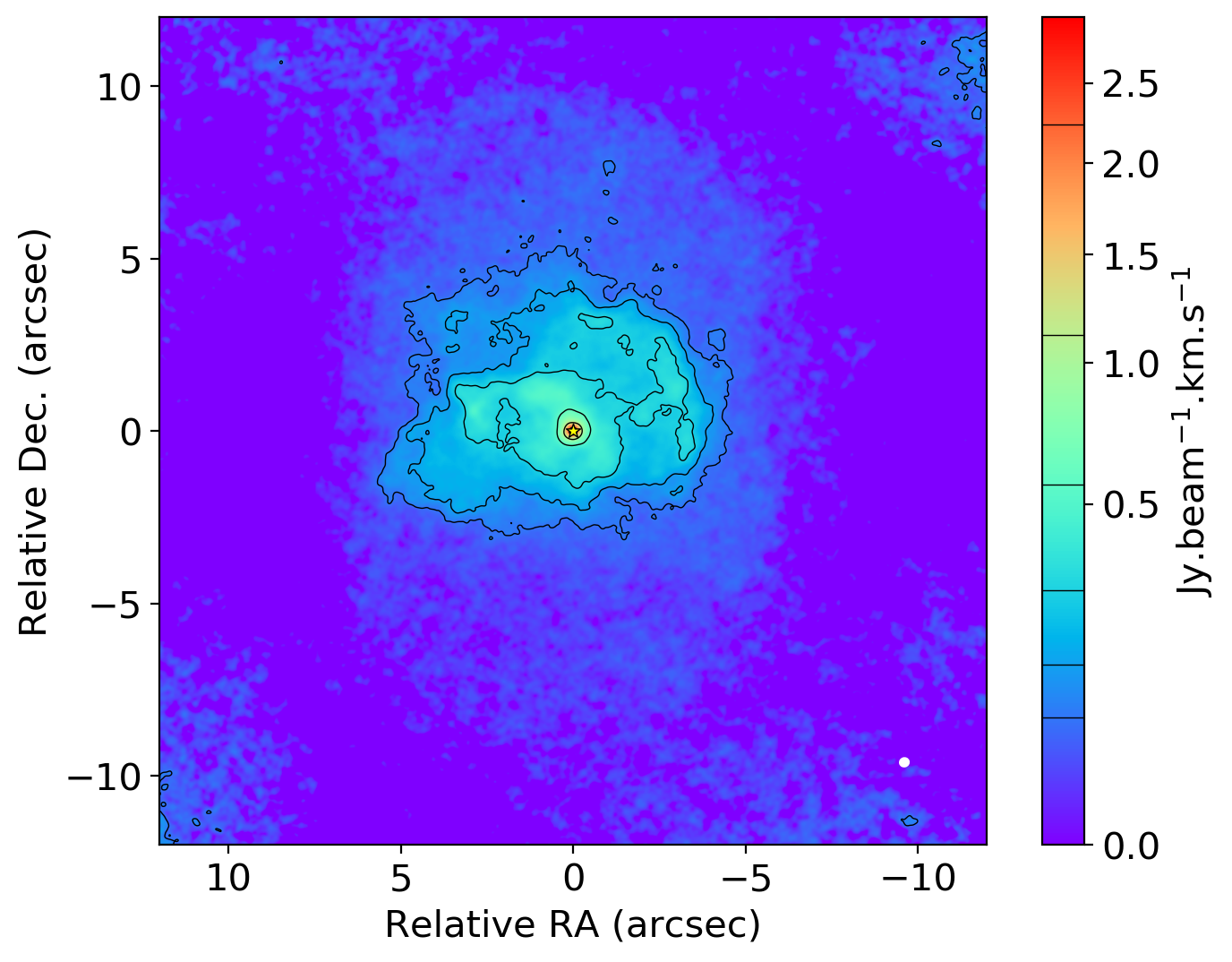}
                \includegraphics[width=8.8cm]{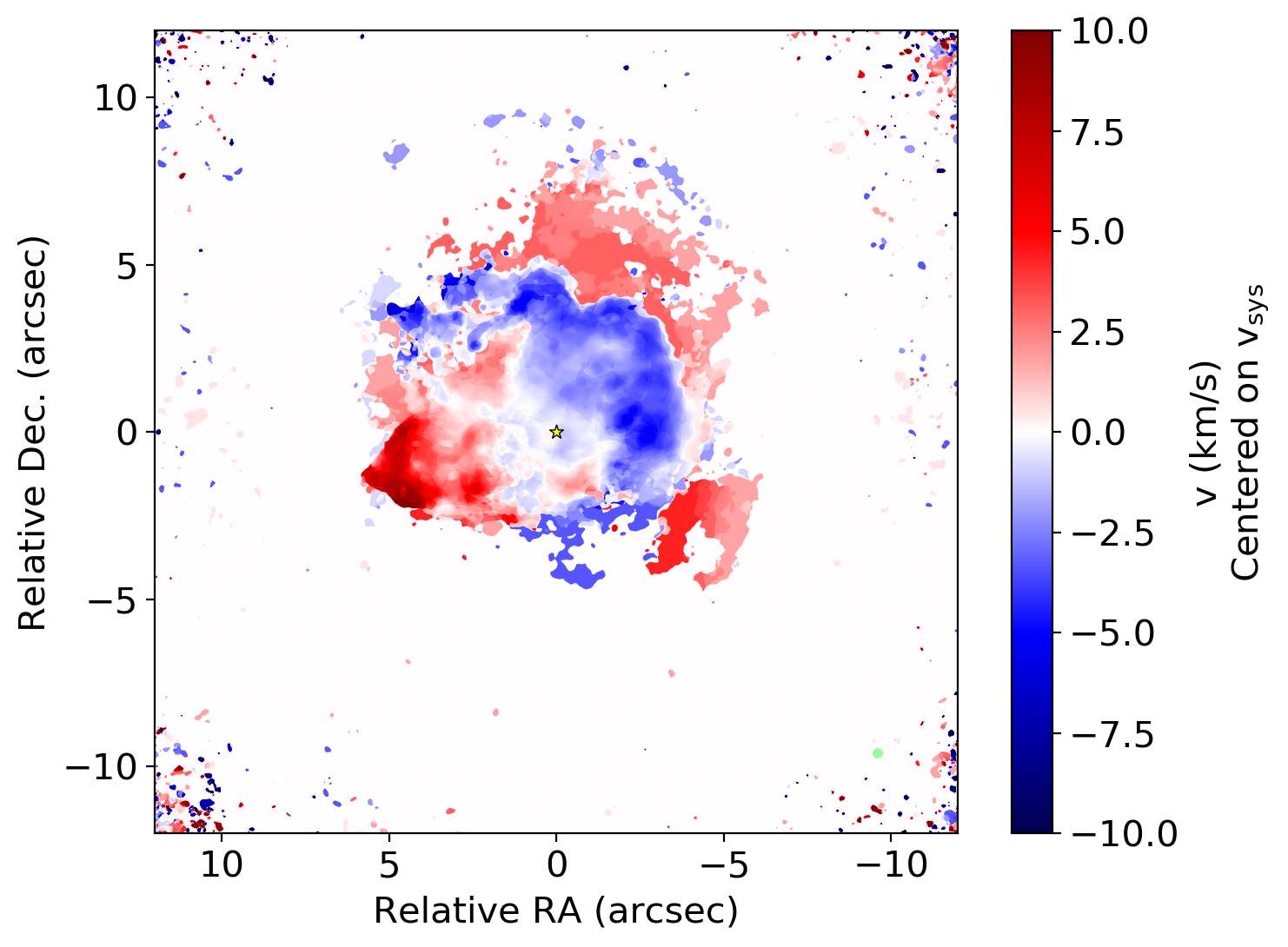}
                \caption{Moment 0 (\emph{left}) and moment 1 (\emph{right}) maps of the LR CO emission in the channel maps shown in Fig. \ref{COchan1}. Contours are drawn at 3, 6, 12, 24, 48, and 96 times the rms noise value in the spectral region of the bandpass without detectable line emission. The continuum peak position is indicated by the yellow star.}
                \label{COmomA}
        \end{figure*}

        \begin{figure}[]
                \centering
                \includegraphics[width=8cm]{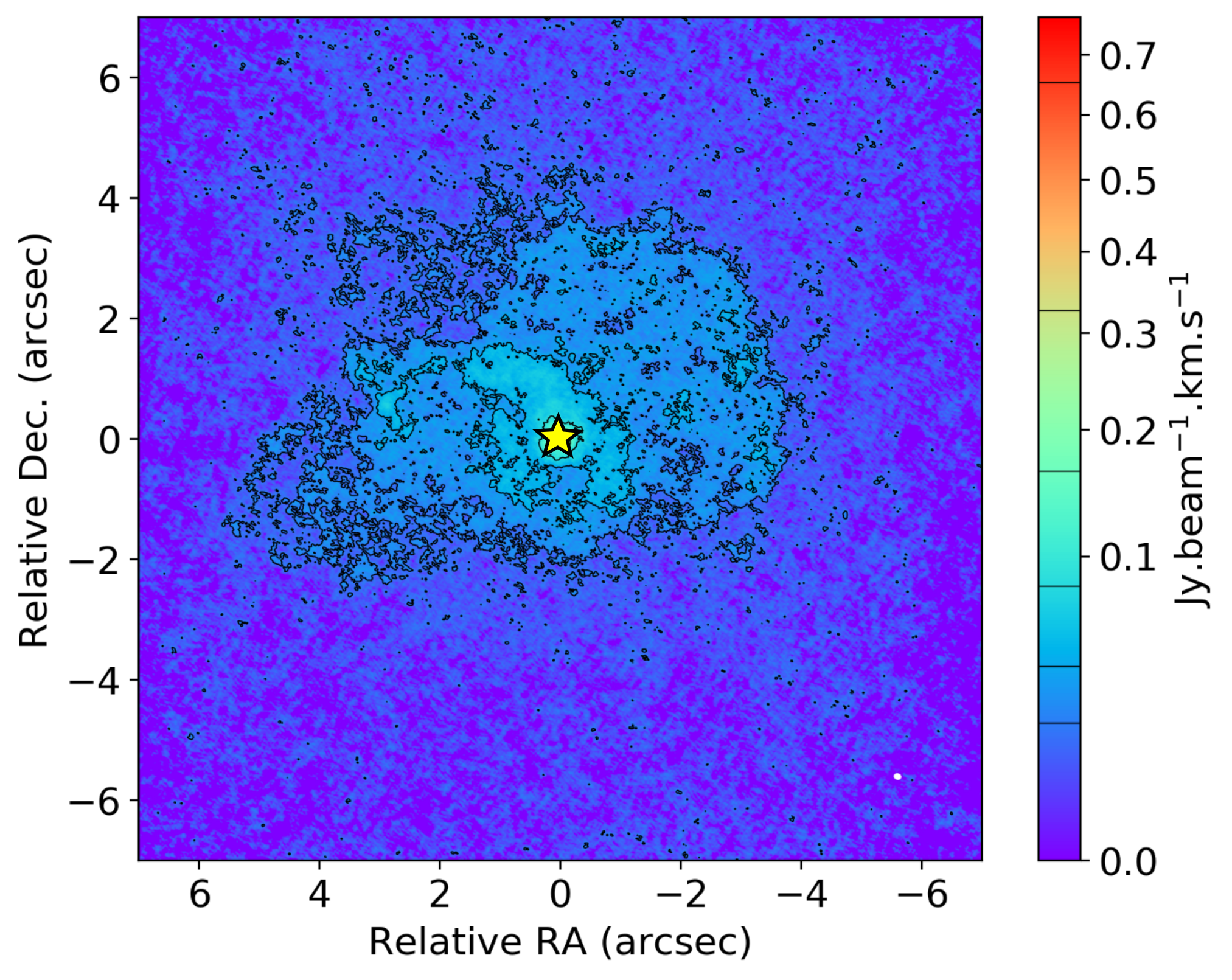}
                \caption{Moment 0 map of the HR CO emission. Contours at 3, 6, 12, 24, 48, and 96 times $\sigma_{\rm rms}$. The continuum peak is at position (0,0), indicated by a yellow star symbol.}
                \label{COmom0}
        \end{figure}

        \begin{figure*}[]
                \centering
                \includegraphics[width=16cm]{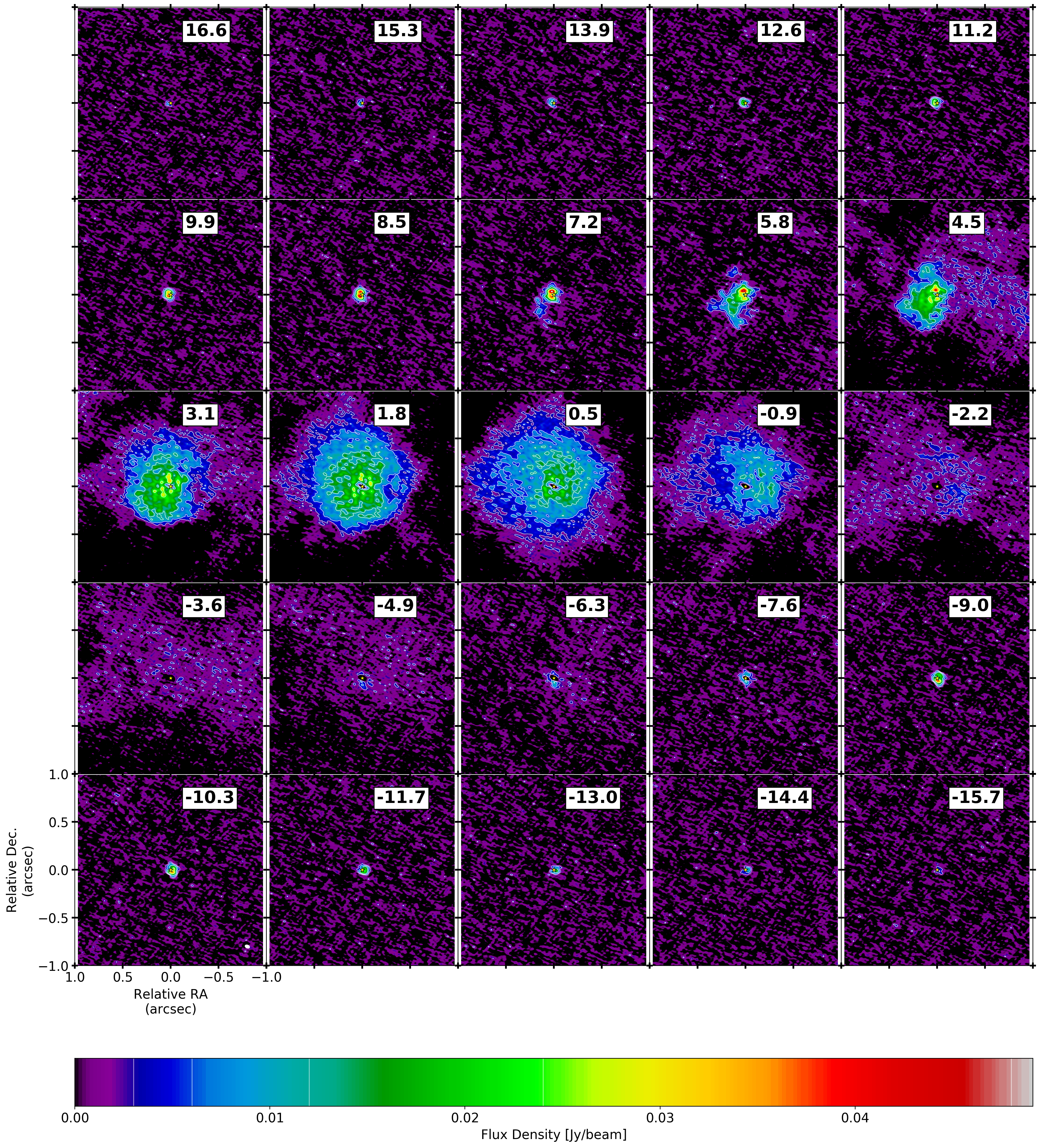}
                \caption{Channel maps showing the spatially resolved emission of the SiO v=0 $J=\ $5$-$4, for the extended dataset. The velocities have been corrected for $v_{*}$=-10.1~$\kms$. Contours are drawn at 3, 6, 12, 24, 48, 96, and 192$\times \sigma_{\rm rms}$ ($\sigma_{\rm rms}$ = 1$\times {\rm 10}^{\rm -3}$ Jy/beam). The ALMA beam has a size of (0.047''$\times$0.03''). Angular scales and are indicated in the bottom left panel. The maps are centred on the continuum peak at position (0,0), indicated by a small yellow star.}
                \label{SiOchanE}
        \end{figure*}

        \begin{figure*}[]
                \centering
                \includegraphics[width=8.5cm]{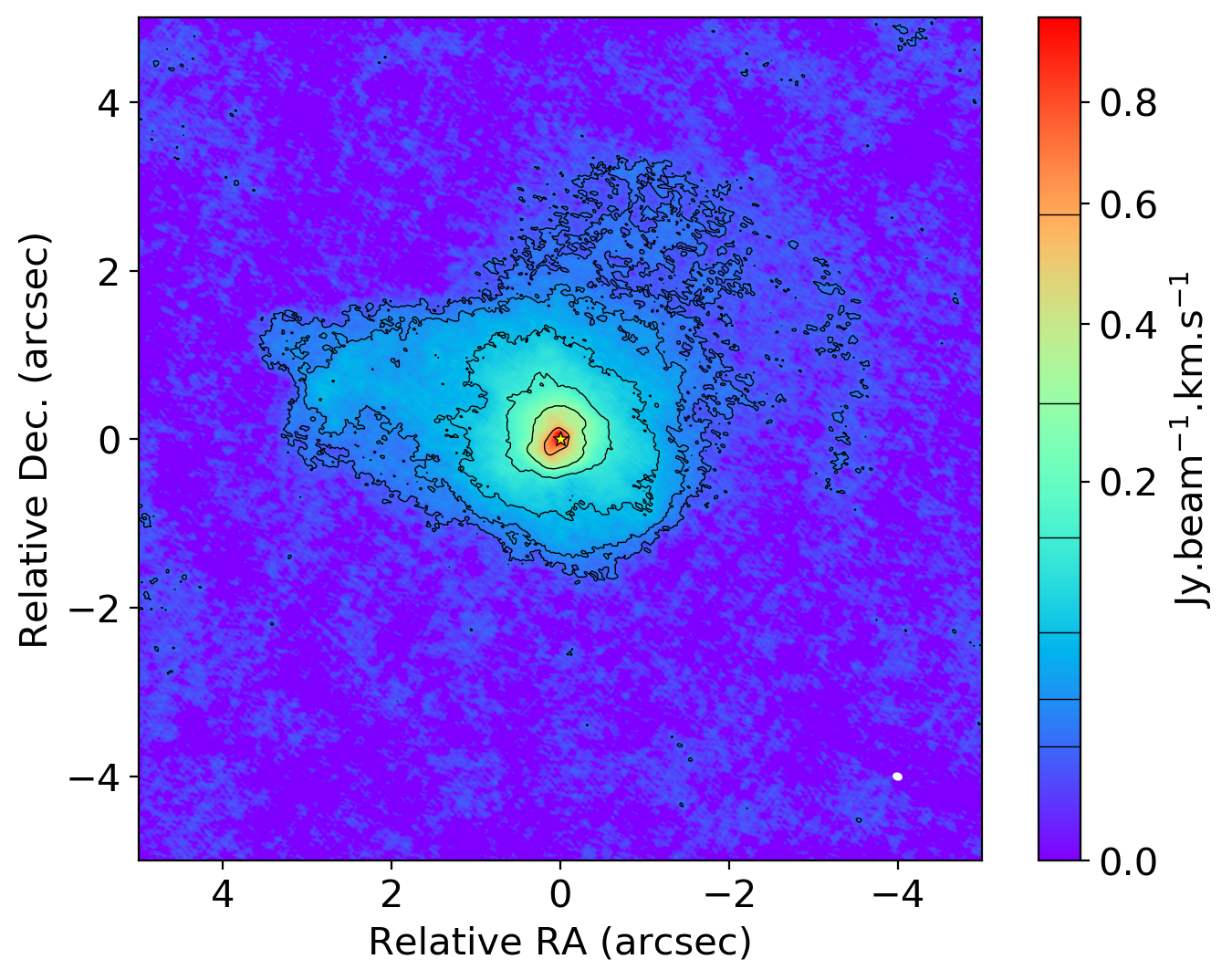}
                \includegraphics[width=8.8cm]{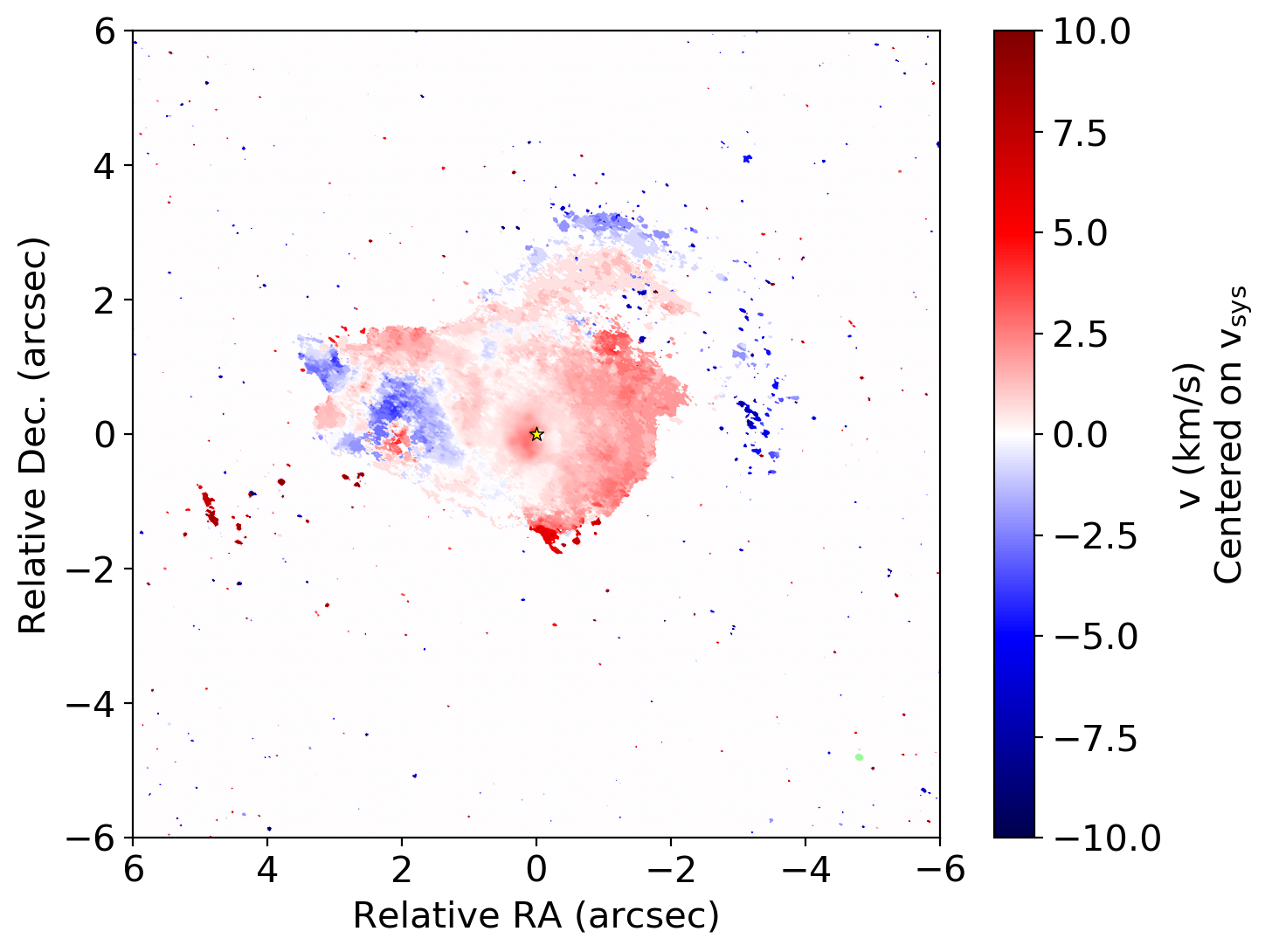}
                \caption{\emph{Left:} Moment 0 map of the SiO v=0 $J=\ $5$-$4 emission. Contours are drawn at 3, 6, 12, 24, 48, 96, and 192 times the rms noise value in the spectral region of the bandpass without detectable line emission ($\sigma_{\rm rms}$ = 8.0$\times {\rm 10}^{\rm -4}$ Jy/beam). \emph{Right:} Moment 1 map of the same emission.}
                \label{SiOmom}
        \end{figure*}

        \begin{figure}[]
                \centering
                \includegraphics[width=9.5cm]{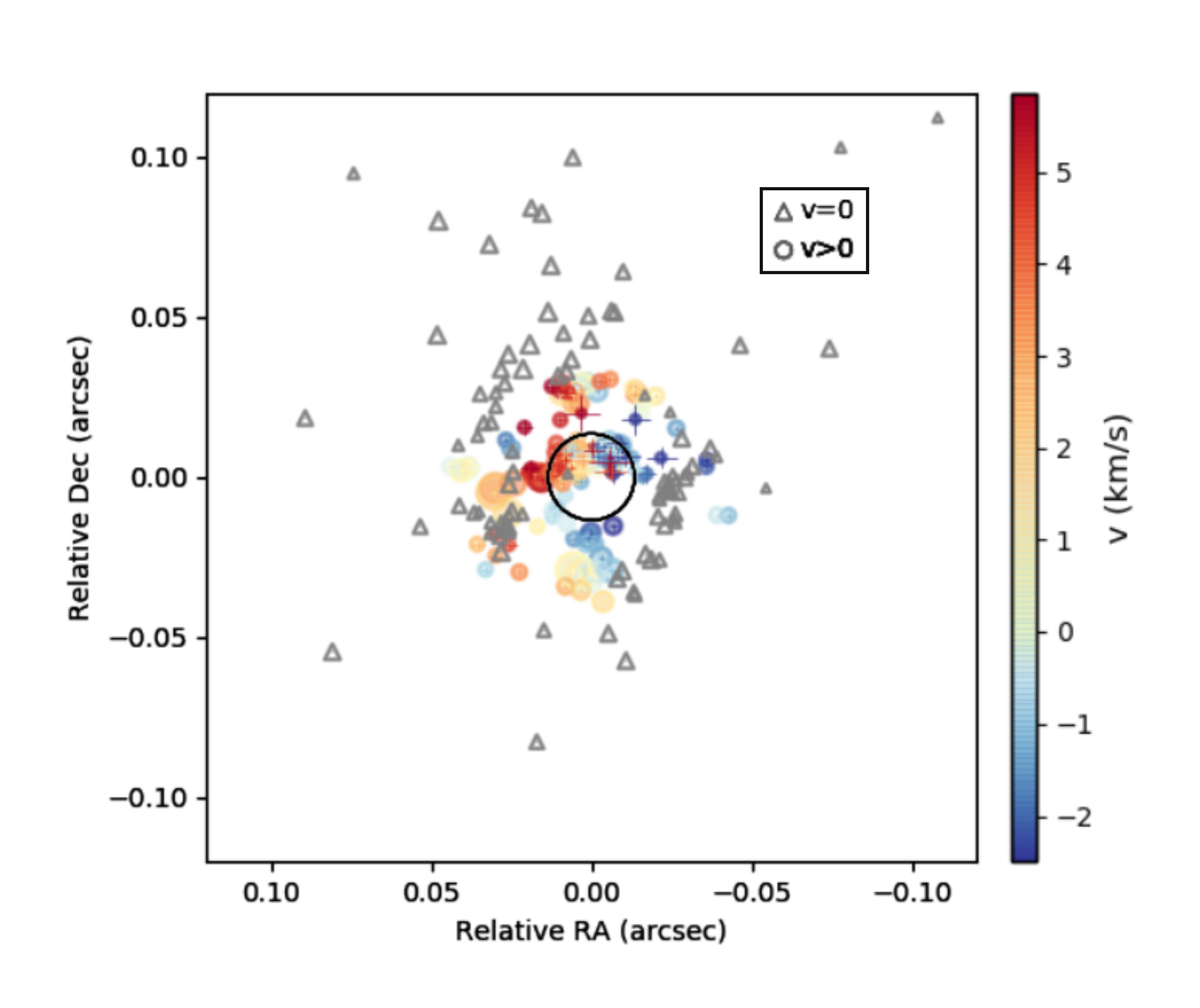}
                \caption{Same as Fig. \ref{masers0E}, but for the higher vibrationally excited lines: $^{\rm 28}$SiO $J=\ $5$-$4 v=1 and 2; $^{\rm 28}$SiO $J=\ $6$-$5 v=1 and 4; $^{\rm 29}$SiO $J=\ $6$-$5 v=1; and $^{\rm 30}$SiO $J=\ $6$-$5 v=1 and 5. The v=0 data are overplotted in grey to better visualise the relative positions of the features.}
                \label{masers1E}
        \end{figure}

        \end{appendix}
        
\end{document}